\documentclass[sigconf]{acmart}
\usepackage{filecontents}
\usepackage{float}

\usepackage{balance}

\usepackage{epsf}
\usepackage{epsfig}
\usepackage{graphicx}
\usepackage{times}
\usepackage{ifthen}
\usepackage{array}
\usepackage{comment}
\usepackage{url}
\usepackage{cellspace}
\usepackage{latexsym}
\usepackage{amsfonts}
\usepackage{subfigure}
\usepackage{multirow}
\usepackage{multicol}
\usepackage{tabularx}
\usepackage{listings}
\usepackage{mathtools}
\usepackage{wrapfig}
\usepackage{xspace}
\usepackage{enumitem}
\usepackage{footnote}
\usepackage{gensymb}
\usepackage{todonotes}
\usepackage{tabularx}
\usepackage{makecell}
\usepackage{xcolor}
\newboolean{showcomments}
\usepackage{pifont}
\usepackage{wasysym}

\usepackage[utf8]{inputenc}
\usepackage{tikz}

\setboolean{showcomments}{true}

\ifthenelse{\boolean{showcomments}}
  {\newcommand{\nb}[2]{
    \fbox{\bfseries\sffamily\scriptsize#1}
    {\sf\small$\blacktriangleright$\textit{#2}$\blacktriangleleft$}
   }
   
  }
  {\newcommand{\nb}[2]{}
   
  }

\newcommand{\add}[1]{{\color{black} {\xspace#1}}}

\newcommand{\tool}{{HomeEndorser}\xspace}
\newcommand{\tools}{{HomeEndorser}'s\xspace}

\newcommand{\no}{{\ding{55}}}
\newcommand{\ye}{\ding{51}}
\newcommand{\pa}{\LEFTcircle}

\newcommand{\ie}{\textit{i.e.},\xspace}
\newcommand{\eg}{\textit{e.g.},\xspace}

\newcommand{\devstate}[1]{{\sf \small {#1}}\xspace}
\newcommand{\home}{{\tt home}\xspace}
\newcommand{\fire}{{\tt fire}\xspace}
\newcommand{\safety}{{\tt safety\_state}\xspace}
\newcommand{\illuminance}{{\tt illuminance}\xspace}
\newcommand{\away}{{\tt away}\xspace}
\newcommand{\secstate}{{\tt security\_state}\xspace}

\newcommand{\nest}{{NEST}\xspace}

\newcommand{\smartthings}{{SmartThings}\xspace}
\newcommand{\homeassistant}{{HomeAssistant}\xspace}

\newcounter{taskctr}

\newcommand{\myparagraph}[1]{\vspace{0.5em}\noindent{\bf #1}:}
\newcommand{\emparagraph}[1]{\vspace{0.5em}\noindent\underline{\em #1}:}

\definecolor{dkgreen}{rgb}{0,0.6,0}
\definecolor{gray}{rgb}{0.5,0.5,0.5}
\definecolor{mauve}{rgb}{0.58,0,0.82}
\definecolor{shadecolor}{rgb}{0.95,0.95,0.95}
\setlist[itemize]{leftmargin=*, noitemsep, topsep=1pt}
\setlist[enumerate]{leftmargin=*, noitemsep, topsep=1pt}

\makeatletter
\renewcommand{\paragraph}{%
	\@startsection{paragraph}{4}%
	{\z@}{0.5ex \@plus 0ex \@minus .2ex}{-1em}%
	{\normalfont\normalsize\bfseries}%
}
\makeatother

\makeatletter
\g@addto@macro\normalsize{%
	\setlength\abovedisplayskip{0pt}
	\setlength\belowdisplayskip{0pt}
	\setlength\abovedisplayshortskip{-10pt}
	\setlength\belowdisplayshortskip{0pt}
}
\makeatother

\DeclareGraphicsExtensions{.pdf,.png}
\graphicspath{{img/}}

\newcolumntype{P}[1]{>{\centering\arraybackslash}p{#1}}

\renewcommand{\refname}{{\hskip 0pt plus 0.1fil minus 0pt} References Cited\hfil}                   %

\includecomment{submission}
\excludecomment{cameraready}

\begin{document}
\title{Towards Practical Integrity in the Smart Home with HomeEndorser}
\author{Kaushal Kafle}
\email{kkafle@cs.wm.edu}
\affiliation{
\institution{William \& Mary}
\city{Williamsburg}
\state{Virginia}
\country{USA}
}

\author{Kirti Jagtap}
\email{ktj35@psu.edu}
\affiliation{
\institution{Penn State University}
\state{Pennsylvania}
\country{USA}
}

\author{Mansoor Ahmed-Rengers}
\email{mansoor.ahmed@cl.cam.ac.uk}
\affiliation{
\institution{University of Cambridge}
\state{Cambridge}
\country{UK}
}

\author{Trent Jaeger}
\email{trj1@psu.edu}
\affiliation{
\institution{Penn State University}
\state{Pennsylvania}
\country{USA}
}

\author{Adwait Nadkarni}
\email{apnadkarni@wm.edu}
\affiliation{
\institution{William \& Mary}
\city{Williamsburg}
\state{Virginia}
\country{USA}
}

\begin{abstract}

Home automation in modern smart home platforms is often facilitated using trigger-action {\em routines}.
While such routines enable flexible automation, they also lead to an instance of the integrity problem in these systems: untrusted third-parties may use platform APIs to modify the abstract home objects (AHOs) that privileged, high-integrity devices such as security cameras rely on (\ie as triggers), thereby transitively attacking them.
As most accesses to AHOs are legitimate, removing the permissions or applying naive information flow controls would not only fail to prevent these problems, but also break useful functionality.
Therefore, this paper proposes the alternate approach of {\em home abstraction endorsement}, which {\em endorses} a proposed change to an AHO by correlating it with certain specific, preceding, environmental changes.  
We present the \tool framework, which provides a policy model for specifying endorsement policies for AHOs as changes in device states, relative to their location, and a platform-based reference monitor for mediating all API requests to change AHOs against those device states.  
We evaluate \tool on the HomeAssistant platform, finding that we can derive over 1000 policy rules for \tool to endorse changes to 6 key AHOs, preventing malice and accidents for less than 10\% overhead for endorsement check microbenchmarks, and with no false alarms under realistic usage scenarios.  
In doing so, \tool lays the first steps towards providing a practical foundation for ensuring that API-induced changes to abstract home objects correlate with the physical realities of the user's environment.

\end{abstract}

\maketitle

\section{Introduction}
\label{sec:intro}

189.4 million households worldwide have at least one smart home device~\cite{iotdata2020}.
This popularity can be explained in part due to the convenience of home automation, wherein smart home devices automatically react to changes in the user's physical environment. %
 For example, the user may configure a security camera to begin recording when they leave home, but turn OFF when they return to preserve their privacy~\cite{mmk+20}.
Such automation is often expressed using trigger-action programs known as {\em routines}, that automatically perform a device action (turning OFF the camera) in response to a trigger (\eg a change from ``away'' to  ``home'').

Routines are often enabled via third-party integrations/services, which automate device-actions by leveraging platform APIs.
Particularly, third-parties use the APIs to modify two distinct types of objects, {\em device states} (\eg the ON/OFF state of a light bulb), and {\em abstract home objects} (AHOs) that are not device-specific (\eg home/away, hereby referred to as the \home AHO).
These AHOs are in fact computed in several ways, often through third-party services (\eg set by the user~\cite{homeawayUserInput}, inferred by querying a combination of device states~\cite{homeawayDeviceStates}, using some other proprietary approach~\cite{homeawayProprietary}, or inferred from other factors such as geolocation~\cite{homeawayLocation}). 
One particularly dangerous attack vector that emerges from this setting is where adversaries gain privileged access to devices {\em indirectly}, by falsifying an AHO that a high-integrity device depends on (via a routine).
For instance, consider a situation wherein an adversary may want to disable the security camera to perform a burglary unnoticed, but may not have direct API access to it. 
As demonstrated by prior work~\cite{kmm+19}, such a camera (e.g., the \nest security camera) can be remotely disabled by compromising a relatively low-integrity integration (\eg a TP Link Kasa app) and using its privilege to change the \home AHO (\ie from ``away'' to ``home''), thus triggering the routine described previously and turning the camera OFF. 
Given the widespread vulnerabilities in third-party integrations~\cite{kmm+19} and the extensive privileges that enable them to modify AHOs~\cite{fjp16,tzl+17}, this type of lateral privilege escalation is of serious concern.

This problem of lateral escalation is, at its core, {\em an integrity problem} analogous to those seen in operating systems: a high-integrity process (here, the security camera) relies on the value of an object (\ie the \home AHO), which can be modified by untrusted parties. 
Thus, smart home platforms must directly address the {\em the lack of strong \underline{integrity} guarantees for AHOs used in automation}. 
This framing deviates from existing literature where proposed mitigations include restricting over-privilege~\cite{rfep18,lck+17,tzl+17} or restricting permission usage based on the runtime context of an API access~\cite{jcw+17}. 
We believe such mitigations may imposes infeasible usability penalties, whilst not addressing the underlying lack of integrity (see Section~\ref{sec:motivation}).

Information flow control (IFC) has often been proposed as a method to ensure the integrity of information consumed by sensitive processes~\cite{bib77,fbf99,mr92,ekv+05,kyb+07,zbkm06}, through labeling of subjects and objects, and dominance checks that regulate flows based on labels~\cite{bib77}.
For example, the IoT platform may mark the \home AHO with a high-integrity label and mark third-party services chosen by the user as low-integrity. 
However, the enforcement of such a policy may block a service chosen by the user to update \home, result in a false denial from the user's perspective (\eg 19/33 \nest integrations from a prior dataset~\cite{kmm+19} would be blocked under such a model).

IFC systems rely on {\em endorsement}~\cite{cecchetti17ccs,kyb+07,zbkm06,zdancewicz02tcs}
to overcome this limitation, allowing trusted programs to change labels of objects to permit flows that would normally violate the labeling.
However, determining the conditions where an endorsement would be allowable is a challenge; indeed, prior work has often avoided directly addressing this problem, and generally facilitates endorsement by assigning the authority to certain {\em trusted} high-integrity processes, thereby delegating the task of determining {\em how to endorse correctly} to the programmer or administrator~\cite{kyb+07,zbkm06,cw87,sjs06}.
However, in our case, smart home users may lack information about dependencies among devices and AHOs to do this correctly. 
So, we ask instead: Is there something else we can rely on to provide endorsement for {\em practical} integrity guarantees?

Yes -- the cyber-physical nature of the smart home provides us with a unique opportunity for practical endorsement, in the form of ground truth observations from devices (\ie device state changes) that can be used to validate proposed changes to AHOs.
For instance, we can endorse the change to the \home AHO (from ``away'' to ``home'') if the door lock was legitimately unlocked (\ie with the correct keycode) recently, since the lock being unlocked represents the home owner's intent and attempt to enter the home.  
Thus, we state the following claim that forms the foundation of this work:
\vspace{0.75em}

 \hspace*{-0.65cm}
\noindent\fbox{%
    \parbox{1\linewidth}{%
Abstract home objects (AHOs) shared among third-party services and devices for the purpose of home automation are inherently tied to a home's physical state. Thus, any state change or modification to an AHO via an API call can be {\em endorsed} using the {\em local context} of the home which consists of a combination of device states.}
}

\vspace{0.75em}
 
\myparagraph{Contributions} We introduce the paradigm of {\em home abstraction endorsement} to validate changes to AHOs initiated by untrusted API calls, and propose the \tool framework to enable it.
\tool does not continuously monitor AHOs, but focuses on API-induced {\em changes} to specific AHOs, and performs a sanity check using {\em policies} that rely on the recent physical state changes in smart home devices. 
 If the check fails, the state change is denied and the user is informed.
 \tools preemptive action prevents future automation based on maliciously changed AHOs.
We make the following contributions in exploring this novel design space:

 \myparagraph{1. Home Abstraction Endorsement} 
We introduce the paradigm of {\em home abstraction endorsement}, which leverages the states of local devices to endorse proposed changes to AHOs.
Our approach makes IFC endorsement practical by exploiting the cyber-physical nature of the smart home, and  significantly deviates 
 from prior IFC systems that enforce integrity guarantees purely based on opaque labels.%
 
\myparagraph{2. The \tool Framework}
We design the \tool framework to provide secure and practical endorsement, consisting of {\sf (1)} a {\em policy model} that allows a unified expression of location-specific device instances within a single policy (\eg endorsing \home via multiple physical entry points), {\sf (2)} a platform-based {\em reference monitor} to enable \tool to mediate all sensitive state changes using these policies, and finally, 
{\sf (3)} a systematic, semi-automated {\em policy specification methodology} for experts, that only considers device-states/attributes that can be trusted for endorsement (\eg read-only, or designated as highly trusted by platforms).

 \myparagraph{3. Implementation and Evaluation} We implement \tool on HomeAssistant, a popular open-source smart home platform, and evaluate it using experimental and empirical analyses.
We first demonstrate that the idea of home abstraction endorsement is feasible, even when using a limited set of trusted device attributes, by generating 1023 policies for endorsing the \home AHO.
We demonstrate the generality of our policy model by identifying several attributes that my be used to endorse 5 additional AHOs. %
We then demonstrate the security guarantee enabled by \tool using a case-study approach.
Moreover, we perform 10 realistic home usage scenarios~\cite{jm17} in a smart home (apartment) testbed to demonstrate that \tool is generally not susceptible to false denials, and in fact, may prevent accidental unsafe situations created through intentional user actions.
We demonstrate \tools practical performance overhead with micro/macro benchmarks (9.7-12.2\% on average).
Finally, we demonstrate that the effort required to specify policies, configure \tool in end-user homes, and integrate \tool in popular smart home platforms, would be modest, and acceptable given the benefit of strong integrity enforcement.

\section{Motivation}
\label{sec:motivation}
\add{
Programmable smart home platforms are popular among users for two primary reasons: {\em integration} and {\em automation}.
That is, platforms such as \smartthings~\cite{smartthingsv3} and \nest~\cite{nestapp} provide permission-protected APIs that enable {\em seamless integration of a diverse array of third-party services}, ranging from software support (\eg cloud and mobile apps) for a variety of smart home devices (\eg TP Link Kasa~\cite{kasastats}, Wemo~\cite{wemostats}), to {\em third-party automation platforms} that hook into the smart home and enable user/developer-provisioned automation routines (\eg~IFTTT~\cite{iftttapp} and Yonomi~\cite{yonomiapp}).

Platform-provided API access allows the caller (\ie the third-party service) to read/write to two broad categories of ``states'' in the home: 
{\sf (1)}~{\em device states}, \ie values associated with individual devices such as the security camera (\eg the ON/OFF state, battery status), and {\sf (2)}~{\em abstract home objects (AHOs)} that are not associated with any specific device, and instead may be {\em computed} in several ways, often by taking into account {\em the user's choice}.
For example, an object that stores whether the user is home or away is an AHO (\ie the \home AHO), and a third-party service designated by the user may compute it in several ways, \ie based on the location of the user's phone~\cite{homeawayLocation}, based on the certain device states in the home~\cite{homeawayDeviceStates}, using a proprietary/undisclosed method/device~\cite{homeawayProprietary}, or through a direct command from the user~\cite{homeawayUserInput}.
The use of AHOs by third-party services, and in routines that control security-sensitive devices in the home, may lead to severe consequences for security and safety, and particularly, the {\em integrity} of the home, as we discuss in the following motivating example inspired from prior attacks~\cite{kmm+19,kmm+20}.
}

\subsection{Motivating Example} 
\label{sec:motivating-example}
\add{

\begin{figure}[t]
	\centering
    \includegraphics[width=3.3in]{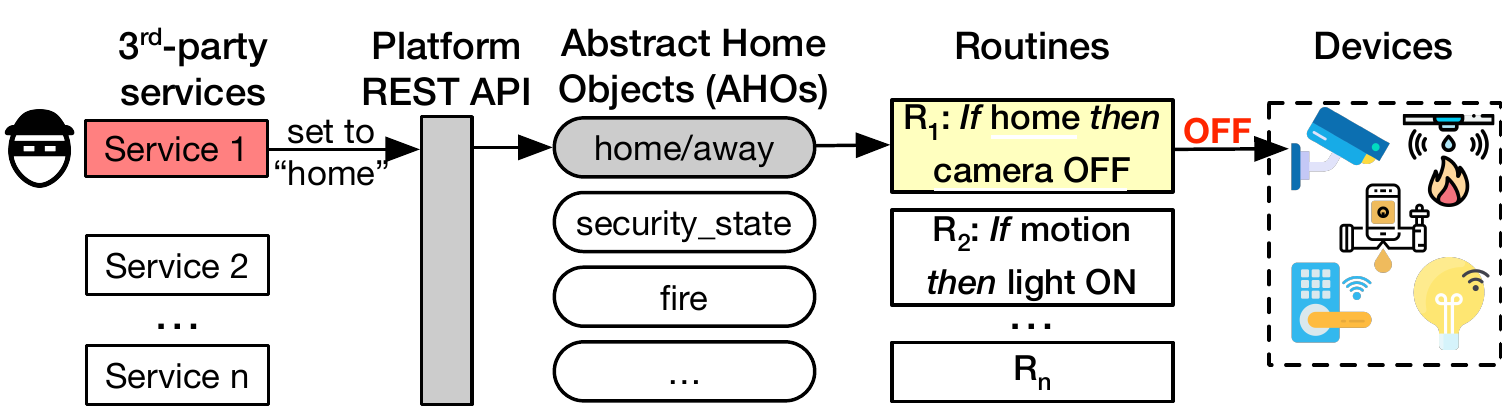}
    \vspace{-1em}
    \caption{\small An attack on the security camera through the manipulation of two shared state objects by adversary-controlled integrations.} 
    \label{fig:attack_graph_example}
    \vspace{-1.5em}
\end{figure}

Alice is a smart home user who has installed a security camera to deter burglars.
For automated monitoring when she is away, Alice has configured two routines for the camera (similar to routines enabled by SimpliSafe~\cite{routinesimplisafe} and \nest~\cite{routinenest}): {\sf (1)} the camera should turn ON when Alice leaves (\ie goes \away), and {\sf (2)} the camera should turn OFF when Alice returns (\ie returns \home).
Additionally, Alice has also integrated several third-party services such as the TP Link Kasa app (which allows her to control her Kasa switch).%

We now introduce our adversary Bob, who seeks to burglarize the house when Alice is away, {\em without being monitored by the camera}.
Bob knows what third-party services are connected to Alice's home, and any vulnerabilities they may have (\eg SSL misuse in the TP Link Kasa app).
Thus, we assume that Bob either exploits one or more vulnerabilities in a connected third-party service to {\em control} it, or steals an authorization token belonging to a service, which allows Bob to access the APIs with the permissions assigned to the service.

Using these abilities, Bob can disable the camera {\em without API access to it} in the manner shown in Figure~\ref{fig:attack_graph_example}: 
Bob uses the access he has obtained from the compromised service (\eg a stolen authorization token from Kasa~\cite{kmm+19}) to set the \home AHO to the value ``home'', falsely suggesting that Alice is home. 
This false change triggers the routine described previously, and turns the security camera OFF.

This problem is not just limited to the \home AHO.
Consider another AHO, the \secstate, which is often used in routines~\cite{routinetotalconnect,routinering} to control security-related devices such as cameras and glass break detectors.
That is, the security devices are armed when the \secstate is set to ``deter'', and disarmed/ignored when the \secstate is set to ``ok'' (similar to routines from Ring~\cite{routinering} and TotalConnect~\cite{routinetotalconnect}).
If Bob can obtain control over an integration that has access to the \secstate, Bob can set it to ``ok'' and disable the camera just as well. 
That is, the adversary may falsify {\em one of many AHOs} to {\em transitively} manipulate a highly sensitive device.

}

\vspace{-1em}
\subsection{The Need for Proactive Integrity Checks}
\label{sec:need-checks}
\add{
As shown in Figure~\ref{fig:attack_graph_example}, Bob transitively attacks a high-integrity device, \ie by modifying AHOs to trigger routines that manipulate a high-security device such as a camera, which Bob {\em cannot compromise or modify (via API) directly}.
In other words, this problem is a smart home-specific instance of the classical {\em OS integrity problem} (\eg Biba integrity~\cite{bib77}), wherein a high-integrity process (\ie the camera) relies on an object (\ie the \home AHO) which can also be modified by low-integrity process (\eg the Kasa integration).

As high-integrity devices rely on such AHOs, traditional wisdom dictates that low-integrity (or third-party) integrations must be disallowed from writing to these objects.
However, recall that in existing platforms (\eg\ \smartthings and \nest), AHOs are often {\em computed} by third-party services chosen by the user, and such integrations may genuinely need to modify AHOs to perform their functionality.
Thus, disallowing {\em the user's choice} of third-party integration from writing to AHOs is bound to break numerous useful services that the user relies on (\eg IFTTT, Yonomi, Kasa), which is a prohibitive cost in terms of user experience that platforms may find undesirable.
For example, in 2019, Google announced an end to its ``Works with NEST'' developer program in favor of a more restricted ``Works with Google Assistant'' program that would only be open to vetted partners~\cite{wwnClosedOriginal}, but immediately backtracked~\cite{wwnClosedFollowup1, wwnClosedFollowup2, wwnClosedFollowup3} given significant opposition from both users and third-party integrations/developers~\cite{wwnClosedFallout1, wwnClosedFallout2, wwnClosedFallout3, wwnClosedFallout4, wwnClosedFallout5}, eventually offering a more flexible program that would allow a broader set of partners access to internal home states (including AHOs)~\cite{wwnClosedFollowup1}.
It is important to note that even when such partners may be vetted, prior work has demonstrated that such vetting is often ineffective, as even vetted integrations may flout security/privacy-design policies~\cite{kmm+19,kmm+20,lsa+21}.
This is particularly true in the case of smart home platforms, as integrations are often hosted {\em outside of the purview of the platform}, \ie on third-party clouds, and any security assessment by the platform given this limited visibility is bound to be only partially effective.

Therefore, there is a need for a solution that is (1) {\em backwards compatible} or practical, \ie does not break functionality by preventing third-parties from accessing AHOs, and (2) {\em effective}, \ie which enables integrity guarantees for devices that rely on AHOs. 
To this end, this paper proposes the moderate route of proactive integrity checking, \ie\ {\em runtime validation of \underline{proposed} changes to AHOs}, which will enable reasonable integrity guarantees while being compatible with platform design and integration choices.
}

\section{Overview}
\label{sec:overview}

\begin{figure}[t]
	\centering
    \includegraphics[width=2.8in]{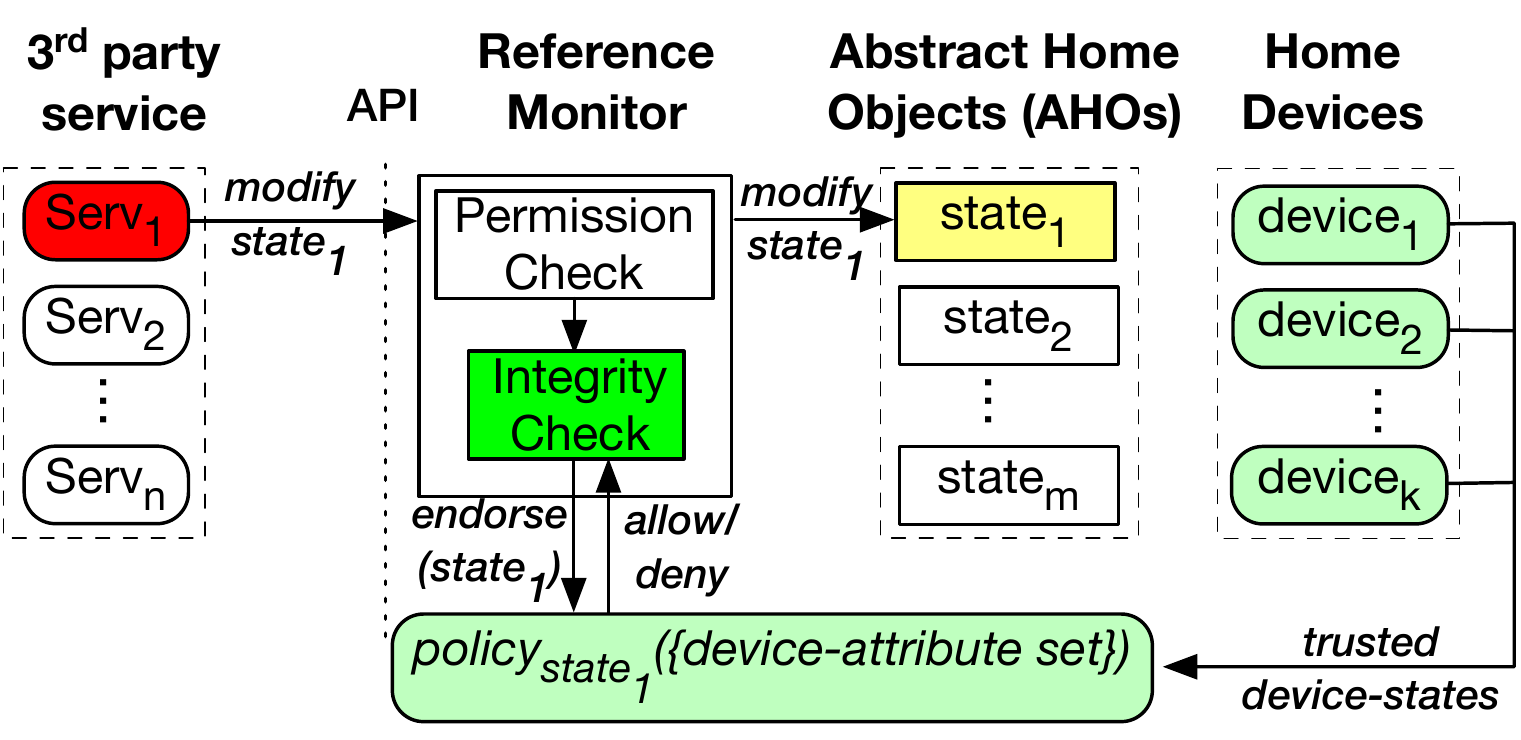}
    \vspace{-1.5em}
    \caption{\small A conceptual overview of physical home endorsement.} 
    \label{fig:overview}
    \vspace{-1.5em}
\end{figure}
\add{
This paper introduces the novel paradigm of {\em home abstraction endorsement} that provides a strong {\em integrity guarantee} for AHOs, defined as follows: In the event that an untrusted service uses the platform API to modify a critical AHO, \eg\ \home or \secstate, the modification will be allowed iff it is consistent with the {\em local state of the home}, composed of the physical device states. %
Our approach builds upon the concept of trusted ``guards'' in the Biba integrity model~\cite{bib77}.
That is, in a system operating under Biba integrity, a high integrity subject cannot receive input from a low integrity subject unless it is endorsed by a trusted guard. 
Similarly, in the smart home, we envision that endorsement policies that apply trusted device states will serve as the trusted guards, ensuring the validity of API requests to change the AHOs that high-integrity devices rely on.

Figure~\ref{fig:overview} provides an overview of our approach.
When a third party service attempts to modify an AHO, the platform first enforces a permission check.
We envision an additional {\em integrity check} for {\em endorsing the proposed change}, performed using an {\em endorsement policy} corresponding to each AHO-change that we plan to endorse.
This policy leverages the recent changes in specific device-attributes/states for arriving at an endorsement decision. 
For example, for endorsing a proposed change in the \home AHO from ``away'' to ``home'', {\em one} possible policy would be as follows: {\em if the door-lock has been unlocked recently (\ie using the correct keycode)}, then \devstate{ALLOW} the change, else DENY.
To build a system that enables our vision of home abstraction endorsement, we first define a set of key design goals, followed by the threat model.
}

\subsection{Design Goals}
\label{sec:goals}
The following goals guide our design of an effective, backwards compatible, framework for home abstraction endorsement:

\begin{enumerate}[label=\textbf{G\arabic*}\xspace,ref=\textbf{G\arabic*}\xspace]\renewcommand{\itemsep}{-0em}
    
    \item \label{goal:policy}\textit{Expressive and Grounded Endorsement Policies.}
 	The endorsement policy structure must be designed in a way that allows it to {\em express common deployment factors} in smart homes that may affect the endorsement, such as {\em device availability} and {\em locality}.
	Moreover, when policies are specified using this design, the specification methodology must be grounded in real smart home devices and their characteristics for practical relevance.%

	\item \label{goal:complete-mediation}\textit{Complete Mediation.} 
	Third party services are generally deployed in cloud environments outside the platform's control, as simple cloud end-points that can make REST API calls (\eg the ``Works with Google Assistant'' platform~\cite{wwGoogleAssistant}, or \smartthings v3~\cite{smartthingsv3}).
	Hence, the reference monitor should be {\em app-agnostic}, \ie should not depend on the analysis/instrumentation of apps/services, but should provide complete mediation 
	for all API calls that modify AHOs, {\em irrespective of app logic}.

 	\item \label{goal:tamperproofness}\textit{Tamperproofness.}
 	While our endorsement approach relies on device states, several device states may be modifiable by untrusted services using the platform API.
 	We need to build a reference monitor that {\em only relies on trusted reports} from devices.

	\item \label{goal:freshness}\textit{Freshness.} 
	The task of endorsing an AHO change may require the reference monitor to examine {\em recent changes} in the states of physical devices, rather than simply reading the current state (\eg as sensor states may reset after apprising the platform of an event).
	Hence, the reference monitor must have trusted access to historical states and timestamps.

	\item \label{goal:performance}\textit{Minimal Performance and Management Overhead.} Implementing physical state endorsement is likely to introduce several additional performance costs to API accesses.
	The framework should minimize any performance delay {\em perceivable by the user}, as well as the effort required to deploy and manage policies.

\end{enumerate}

\subsection{Threat Model}
\label{sec:threat_model}
\add{
We consider a network-based adversary with the ability to compromise any third-party integration/service connected to the target's home (\eg as demonstrated by prior work~\cite{kmm+19,fjp16}). 
The attacker's objective is to {\em indirectly} modify or disable high integrity devices (\eg security camera) using compromised or malicious third-party services.
The adversary may issue API commands to the target's home that may affect device states; however, the adversary does not have the ability to {\em compromise devices} in the home by other means (\ie which is a standard assumption in work that deals with API misuse~\cite{ctm19,dh18,dhc21}).
The adversary cannot compromise the platform and the platform app; explicitly, the adversary cannot censor notifications sent from the platform to the user.
Devices may be offline; however, we do not account for byzantine fault tolerance, but point to complementary prior work~\cite{bem19} that verifies reported device states.
}

\section{The \tool Framework}
\label{sec:design}

We propose \tool, a framework that enables home abstraction endorsement by developing the methods and systems necessary to achieve the design goals~\ref{goal:policy}$\rightarrow$\ref{goal:performance}. 
We first define a {\em policy model} that can express deployment-considerations such as device locality and availability (\ref{goal:policy}).
\tool automatically instantiates the policies defined within this model in the context of the end-user's home, i.e., considering the devices available as well as install locations, thereby reducing deployment-time configuration effort (\ref{goal:performance}).

\tools\ {\em reference monitor} is integrated into the user's smart home platform in the form of an endorsement check in the platform subsystem responsible for executing all API calls, thereby ensuring complete mediation (\ref{goal:complete-mediation}).
When a third-party makes an API call to modify an AHO-change that \tool endorses, \tool invokes the corresponding policy, and uses a {\em state machine} to retrieve the {\em most recent change} in each relevant {\em device state/attribute} included in the policy~(\ref{goal:freshness}).
We consider the most recent change, rather than the current state of the device attribute, as the two may be different (since most sensors reset after a change), and because the most recent device state changes provide the context for endorsing the proposed AHO change.
This design decision is instrumental in eliminating unnecessary false denials (see Section~\ref{sec:eval-intentional-changes}).

We formulate a {\em policy specification methodology} that allows experts to specify endorsement policies in a systematic, ground-up manner (\ref{goal:policy}).
Our approach derives a mapping of smart home devices and their {\em attributes} (\ie fields that indicate specific device states), which are further filtered to a set of {\em endorsement attributes}:  device-attributes that can be {\em trusted} for endorsement, due to being either read-only or treated as highly restricted by platforms (\ref{goal:tamperproofness}).
Once the trusted attributes are identified, we use systematic open coding to identify {\em inferences} drawn from them that can endorse specific AHOs.
These inferences, when expressed in our policy model, form deployment-independent {\em endorsement policy templates} that are automatically instantiated in the user's deployment-context.

The rest of this section describes the three design components, starting with the policy model (Section~\ref{sec:policy-design}), followed by the runtime enforcement (Section~\ref{sec:endorsement-check}), and the data-driven policy specification and instantiation methodology (Section~\ref{sec:policy-spec}).
We use the endorsement scenario in Section~\ref{sec:motivating-example} as the running example.%

\subsection{Policy Model}
\label{sec:policy-design}

A key challenge for \tool is designing a policy model that can alleviate two  practical constraints.
First, unlike our motivating example where a single device-attribute (\ie the \devstate{UNLOCKED} state of the door lock) was used for endorsing \home, in practice, endorsement policies may consist of more than one device-attribute that must be checked simultaneously.
Second, the device location is an important factor in deciding what they can endorse.
For example, while a door lock on the front door as well as a door lock on the back door can both indicate that the user is \home, the two endorsements are naturally {\em mutually exclusive}.
We account for these constraints with a policy design that can be expressed as a Disjunctive Normal Form (DNF) boolean formula, as follows:

\vspace{1em}\noindent {\bf Definition 1} (Endorsement Policy). The policy for endorsing a change in AHO $x$ to value $y$, $P_x(y)$, is a DNF formula composed of one or more location-specific predicates ($L_i$), \ie 
$P_x(y) = L_1$ $\vee$ $L_2$ $\vee$ $...$ $\vee$ $L_n$, where a location-specific predicate is defined as follows:

\vspace{1em}\noindent {\bf Definition 2} (Location-specific Predicate). A location-specific policy predicate $L_i$ for location $i$ (\eg entryway), \ie $L_i = d_j \wedge d_k \wedge ... d_m$, is a conjunction of one or more device-attribute checks $d_j$, defined as follows:

\vspace{1em}\noindent {\bf Definition 3} (Device-attribute Check). A device-attribute check $d_j$ is a condition $d_j == s$, where $s$ is a physical state that the particular device-attribute must have exhibited in the recent past, for the device-attribute check to return {\em true}. 

To illustrate, let us express the policy from the motivating example for endorsing the \home AHO's change to the value ``home''. We express the policy using a door lock and a motion sensor at the entry way, as well as the same devices at the rear entrance, as follows:
\begin{center}
{\small
$P_{\home}${\footnotesize (home)} $=$ (door-lock\_{lock} $==$ {\small\tt UNLOCKED} $\wedge$\\
			  motion\_{sensor} $==$ {\small \tt ACTIVE})$_{front-door}$

$\vee$ 
	
	(door-lock\_{lock} $==$ {\small\tt UNLOCKED} $\wedge$\\ motion\_{sensor} $==$ {\small \tt ACTIVE})$_{back-door}$
}
\end{center}
The above policy considers both the door lock being unlocked, and motion being sensed, to prevent false negatives.
That is, for both the conditions above to be true, a user would have to unlock the door {\em and then enter}, \ie proving that they are home beyond doubt. 
On the contrary, if the user unlocks but leaves without entering, this policy condition would correctly result in a denial (as shown in Section~\ref{sec:eval-intentional-changes}).
Similarly, the disjunction between the location-specific predicates enables their independent evaluation, thereby allowing the AHO-change as long as either one evaluates to a {\em true} result.
Finally, we define two policy {\em actions}: {\sf \small ALLOW} and {\sf \small DENY}, corresponding to the {\em true} or {\em false} values that the DNF formula results in, respectively.

For Bob to circumvent \tools defense, Bob could attempt to modify \home at the very moment when Alice is performing what are semantically completely opposite actions, but sufficient for the endorsement of \home.
For instance, Alice could be {\em leaving}, which would also involve {\sf (1)} unlocking the door, and {\sf (2)} triggering the door-way motion sensor.
A naive implementation of our policy model would be susceptible to this attack, as it would consider the above device state changes, even if performed in the opposite sense, to be evidence that Alice is returning home, because it matches $P_{home}${\footnotesize (home)}.
However, smart home devices provide {\em unique} device attribute values even for similar actions, \ie the state value for unlocking the door using the keypad is different relative to simply unlocking it from the inside (\ie ``owner'' in the former case, and ``manual'' in the latter).
Our policy specification~\ref{sec:policy-spec} and enforcement (Section~\ref{sec:endorsement-check}) consider device-attribute values at this precise granularity, preventing such an attack.

\subsection{Secure and Practical Enforcement}
\label{sec:endorsement-check}

\begin{figure}[t]
	\centering
    \includegraphics[width=2.7in]{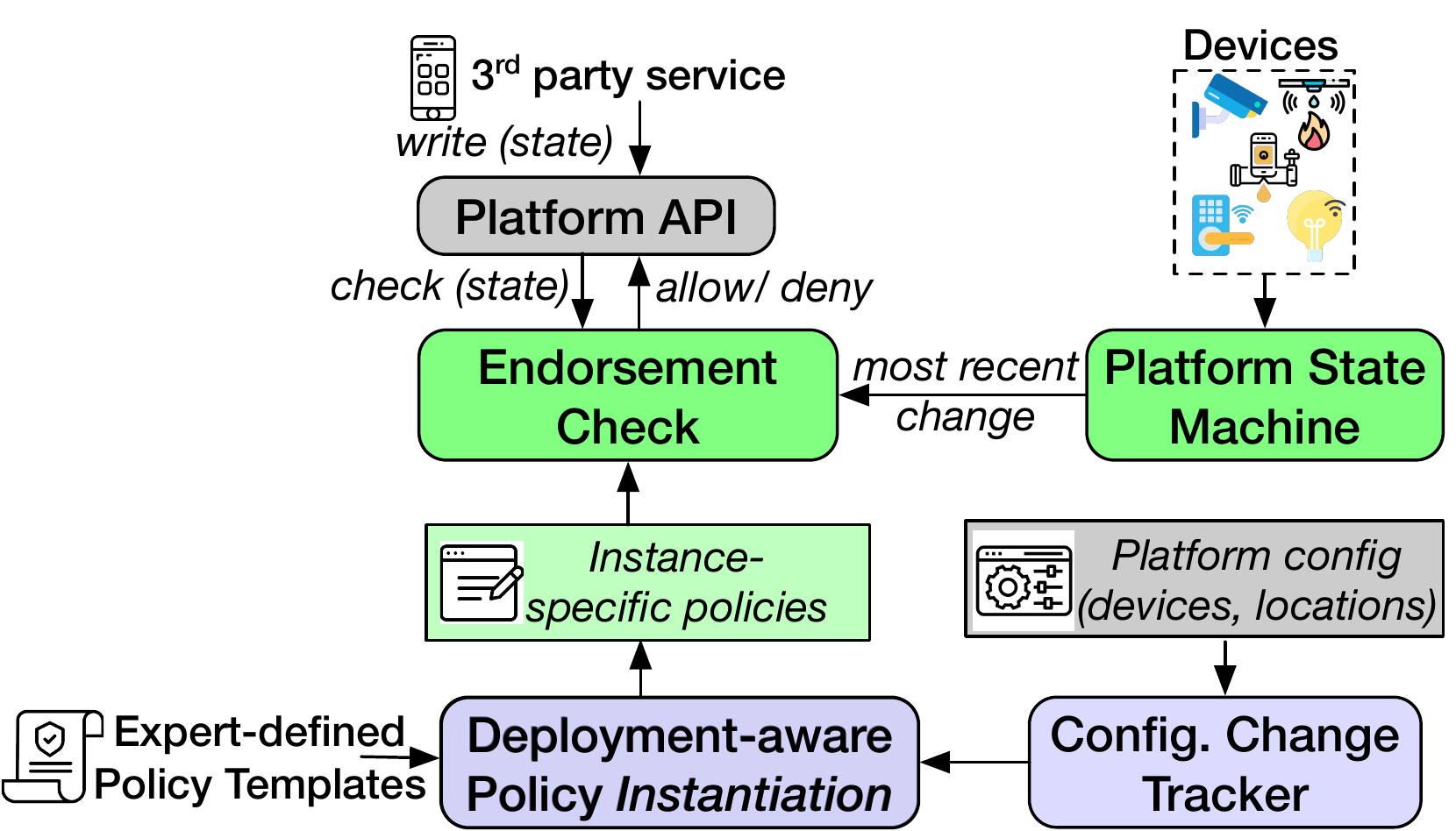}
    \vspace{-1.2em}
    \caption{\small Design components of \tools enforcement} 
    \label{fig:enforcement}
    \vspace{-1.8em}
\end{figure}

We integrate \tools enforcement into the smart home platform, to enable it to mediate all API commands from third-parties before they are executed (\ref{goal:complete-mediation}).
This decision is influenced by how third-party services are currently integrated, \ie as cloud endpoints that use RESTful APIs to interact with the platform, but execute on their own proprietary servers, without a way for the platform to inspect them (\eg\ \nest and \smartthings v3).
Therefore, our decision ensures that the endorsement check will occur regardless of how the integration is implemented, or where it is deployed. 
We now describe our platform-level enforcement in terms of its three design components, as shown in Figure~\ref{fig:enforcement}.

\myparagraph{1. Deployment-aware Policy Instantiation} \tool enforces policies specified as per the policy model described previously (we describe the specification process in Section~\ref{sec:policy-spec}).
However, it cannot directly enforce the general policies without first contextualizing them to a user's home, for three key reasons: {\sf (1)} A typical user's setup is unlikely to have {\em all} the devices specified in the policy, \ie the {\em most restrictive policy} consisting of the most number of devices would always deny the state change. 
{\sf (2)}~The policies need to be configured as per the different locations in which devices are placed in the home, and
{\sf (3)}~The enforced policies need to adapt as the user's smart home evolves (\eg through device addition/removal), to provide the enforcement applicable to the current setup.
Thus, we treat the specified policies as templates that \tool automatically {\em instantiates} in the deployment-context of a particular home.

To elaborate,  when \tool is first set up in a home, it leverages the platform's internal bookkeeping systems to extract all the devices and device-locations.
Then, for each state the user decides to endorse, it uses the policy templates specified previously to instantiate the {\em most restrictive but feasible policy}, \ie the policy that contains the largest aggregate of device-attributes (with the assumption that including more device-attributes would make the inference stronger), but which is also supported given the devices present in the home and their locality.
This most restrictive policy is then marked for enforcement.
Finally, \tool also updates its policy instances upon a configuration change, such as the addition, removal, or relocation of a new device.

\myparagraph{2. The Endorsement Check} \tool mediates all API requests, but only invokes the endorsement check if one of the AHOs selected by the user for endorsement is about to be modified, in a manner similar to performance-preserving {\em hook activations} as previously proposed for Android~\cite{hnes14}~(\ref{goal:performance}).
The check consists of retrieving the policy for the specific AHO-change being endorsed and collecting state information from the device-attributes in the policy.
As described previously in Section~\ref{sec:policy-design}, if the policy decision is {\sf \small ALLOW}, then the proposed change is allowed to go through; else, it is denied, and the user is notified.
A key component of this check is \tools\ {\em platform state machine} that retrieves the device-attribute values, which we now describe in detail.

\myparagraph{3. Retrieving the {\em most recent} changes using the Platform State Machine} 
A naive approach of executing an endorsement check would be to query each device for its {\em current state} at the time of endorsement.
However, such a check would most certainly fail and lead to a false denial because most sensors detect and report a change, and then {\em reset to a predefined neutral state}.
For example, recall the endorsement policy predicate to endorse \home consisting of the door lock and the motion detector (assuming one location for simplicity):
\begin{center}
{\small
	door-lock\_{lock} $==$ {\small\tt UNLOCKED} $\wedge$
			  motion\_{sensor} $==$ {\small \tt ACTIVE}
}
\end{center}

Unless the endorsement check happens exactly at the moment the user enters, the motion detector will reset to its {\sf \small INACTIVE} state immediately after detecting motion, resulting in a false denial.
Therefore, for correct endorsement, we check the {\em most recent but fresh change} in the physical device states (\ref{goal:freshness}), \ie
the last state change before the state automatically reset, within a certain configurable time threshold to ensure freshness (\eg one minute). 

A state machine that keeps track of the state changes in all the devices provides \tools reference monitor with this data, by tracking changes through callbacks placed in the entities that represent devices on the platform (\eg device handlers in \smartthings, or device-integrations in \homeassistant).
Every time the physical state of a device changes, the state machine callback is invoked to store the most recent state change along with the timestamp.
The timestamp helps \tool discard states that are older than the preconfigured threshold, thereby preventing old states from causing {\em false allows} (\ref{goal:freshness}).
Finally, we note that this state machine abstraction does not have to be developed from scratch, and can be built on top of the platform's subsystem that keeps track of device states.

\subsection{Data-driven Policy Specification}
\label{sec:policy-spec}

\begin{figure}[t]
	\centering
    \includegraphics[width=2in]{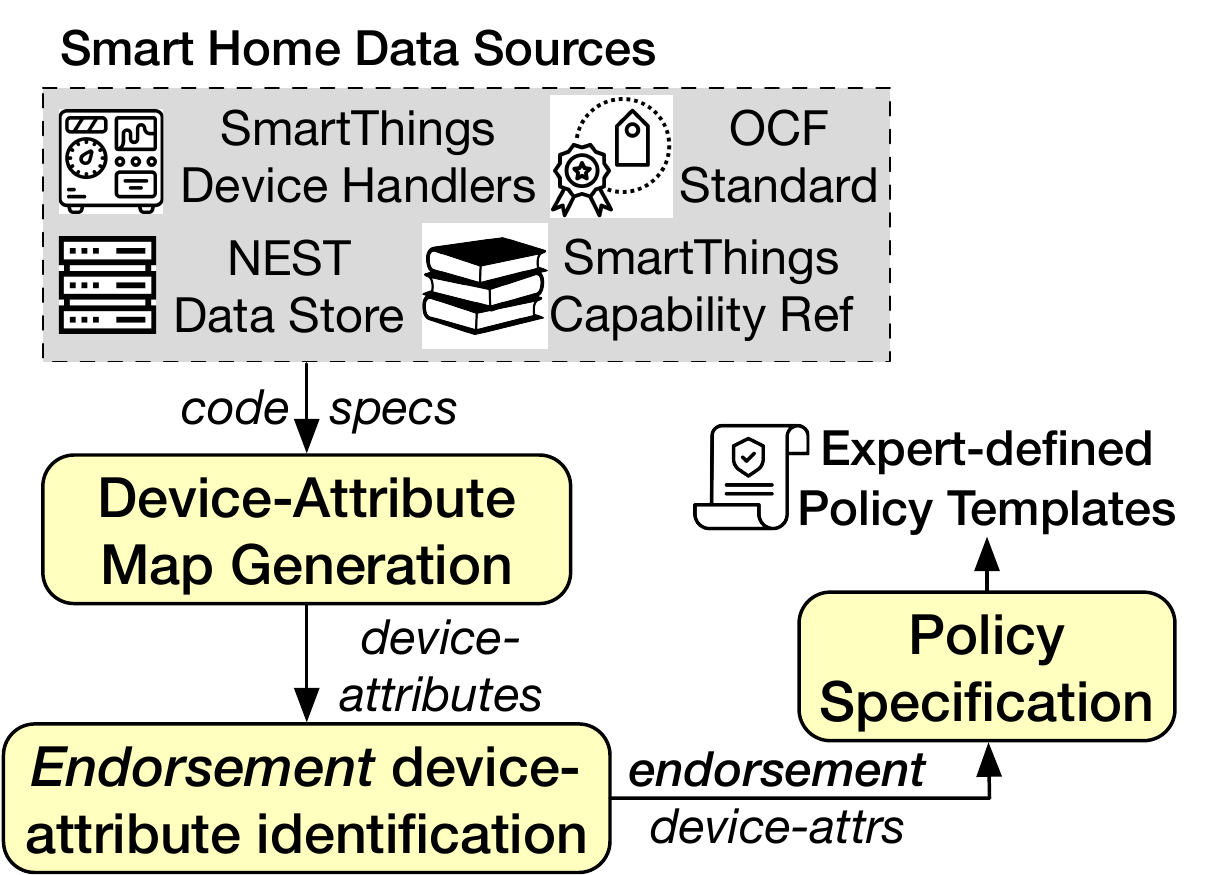}
    \vspace{-1em}
    \caption{\small \tools policy specification methodology.} 
    \label{fig:specification}
    \vspace{-2em}
\end{figure}

Figure~\ref{fig:specification} provides an overview of our data-driven methodology to allow {\em experts} (\eg security researchers, platform vendors) to specify endorsement policy {\em templates} that  are representative of what real device states can enable (\ref{goal:policy}), which \tool automatically instantiates in the context of end-user homes (as previously described in Section~\ref{sec:endorsement-check}).
We begin by creating a {\em device-attribute map}, \ie a comprehensive mapping between device types (\eg cameras, door locks) and the attributes they embody. 
We then define the {\em endorsement attributes} to be used for tamperproof endorsement, and describe our approach for identifying them (\ref{goal:tamperproofness}).
Finally, we use an open coding methodology for identifying the {\em observations} and {\em inferences} that can be made from the endorsement attributes, which are then used to specify policy templates (using the model from Section~\ref{sec:policy-design}).
Note that although we develop a methodology to specify policies, we assume optimal device placement and sensor configuration to be out of scope, and direct the reader to complementary related work that informs users on optimal deployment~\cite{jm17}. 

\myparagraph{1. Generating the Device-Attribute Map}
We develop a semi-automated methodology to create a realistic device-attribute map from platform documentation and
existing device-abstractions such as \smartthings\ ``device handlers''.
We select the following information sources for building the map, based on platform popularity, and the potential of obtaining realistic mappings:
{\sf (1)} a device-resource map from the Open Connectivity Foundation (OCF) specification~\cite{ocfdocs}, used by the open source platform IoTivity~\cite{iotivity}, {\sf (2)} the \nest data store~\cite{nestdoc}, {\sf (3)} the \smartthings capability reference~\cite{smartthingsCaps}, and {\sf (4)} \smartthings device handlers~\cite{smartapp-repo}.
As each of these sources exhibits a unique representation of  devices and attributes, we develop customized, automated, methods for extracting device-attributes from each source. 
To elaborate, OCF explicitly specifies a device types, the attributes associated with each device type in a JSON document~\cite{ocfjson}, which we automatically analyze to obtain a device-attribute map for OCF.
Similarly, the \nest data store provides a mapping of device-attributes in JSON-like format, which we similarly query. %
SmartThings provides a capability reference~\cite{smartthingsCaps}, which lists attributes (\ie capabilities) %
that devices may choose to exhibit, via developer-defined {\em device handlers}, which are software proxies for devices that facilitate device-interaction with the platform.
To extract a device-attribute map from SmartThings, we automatically analyze the 334 device handlers from the \smartthings repository~\cite{smartapp-repo}, and capture the device-attribute relationships evident in the preamble (seen in Figure~\ref{fig:devicehandler} in the Appendix).

\myparagraph{2. Identifying Trusted Endorsement Attributes for Tamperproof Endorsement}
We observe that similar to AHOs, several device-attributes are modifiable by third-parties through the platform APIs, either for legitimate purposes or due to over-privilege. 
Therefore, for ensuring tamperproof endorsement, \tool must be able to trust the information received from the participating endorsers \ie device-attribute pairs (\ref{goal:tamperproofness}). 
We achieve this goal by defining a trusted subset of device-attributes to be used for our checks, \ie\ {\em endorsement attributes}.
We propose two categories of endorsement attributes: {\sf (1)} {\em read-only attributes}, \ie which are only writable by devices, and not via API calls, rendering them read-only from the third-party API caller's perspective (\eg motion sensor reading), and {\sf (2)} {\em designated attributes}, which are writeable in theory, but are considered high-integrity by platforms and prior security research~\cite{sab+17,csk+20} alike (\eg locking the door lock), and hence, heavily restricted.
For example, \nest allows its own platform app to unlock the locks integrated with it (\eg the Nest X Yale Lock~\cite{nestxyale}), but does not allow third-party apps to do the same.
Therefore, both {\em read-only} and {\em designated} device attributes would have a higher integrity level than a state such as \home, and hence, would be trusted to endorse it.

We identify read-only device-attributes in a manner similar to our approach for extracting the device-attribute map, \ie by parsing platform documentation and device handler code.
For identifying designated attributes, we leverage the rationale of popular platforms for assigning integrity levels to attributes only writable by the platform app, particularly those belonging to security-sensitive devices.

\begin{table}
\centering
\footnotesize
\caption{{\small AHOs inferred from Endorsement Attributes}}
\vspace{-1.5em}
\begin{tabular}{c|c}
 \Xhline{2\arrayrulewidth}
\hline
\textbf{AHO} & \textbf{Endorsement attributes}\\
 \Xhline{2\arrayrulewidth}                                                                                                                                                             
\home & $<${\sf security-panel, disarmed}$>$                                                                                       \\ \hline
\home & $<${\sf motion-sensor, active}$>$
	\\ \hline
\fire & $<${\sf temperature-sensor, temperature}$>$
\\ \hline
\fire & $<${\sf smoke-detector, smoke-alarm-state}$>$
\\ \hline
\safety & $<${\sf co-detector,co-alarm-state}$>$
\\ \hline
\illuminance & $<${\sf blind,openLevel}$>$
\\ \hline
\end{tabular}

\label{tab:inference}
\vspace{-1.7em} 
\end{table}
\myparagraph{3. Generating Endorsement Policies from Attribute Inferences}
We now address the question of {\em how} the endorsement attributes are used to endorse a specific AHO, by designing a holistic {\em inference study}.
We begin by identifying 5 additional AHOs, by considering the objects that have been previously examined and found to be susceptible to transitive attacks~\cite{dh18,kmm+19}, and by considering AHOs not mentioned in prior work, but which we encountered when building our device-attribute map (\eg the \secstate from \nest, which prior work does not account for).
Then, we consider each device-attribute, and identify the type of information sensed or observed by that device-attribute, which we then translate to an inference that could be used for endorsing an integrity-sensitive change in one or more of the AHOs.
For example, the device-attribute pair \devstate{$<$security-panel, disarmed$>$}  indicates that the security panel/keypad was recently disarmed, which may indicate that the user recently arrived home, and hence, provide an inference to endorse the \home AHO's proposed change to ``home''.
As a key principle in the construction of policies is the a combination of device inferences, we combine inferences to construct deployment-independent policy templates using the structure defined in Section~\ref{sec:policy-design}.
An example set of inferences discovered using this process are shown in Table~\ref{tab:inference}.

\section{Implementation}
\label{sec:implementation}
This section describes our policy specification study, as well as the reference monitor implemented in \homeassistant.

\myparagraph{1. Policy Specification Study} 
We automatically generated a combined device-attribute map from all the data sources consisting of 100 device-types and 510 device-attribute pairs. 
Of these, we identified 41 endorsement attributes, \ie read-only or designated device-attributes.
Two authors then independently identified the inferences that could be drawn from these endorsement attributes to endorse changes in one or more of our 6 AHOs. 
When identifying inferences, the coders disagreed on 12 out of the 510 device-attribute pairs (2.4\% disagreement rate), which were resolved through mutual discussion.
The inferences specified for these endorsement attributes led to 1023 policies for the \home AHO alone.
Note that the generation of the device-attribute map was fully automated and directly derived from platform sources, and hence, did not result in any disagreements.%
Our data is available in our online appendix~\cite{onlineappendix}.

\myparagraph{2. Implementation on HomeAssistant}
We implemented \tool in HomeAssistant, a popular open-source smart home platform, for two main reasons: {\sf (1)}  the availability of source code, and {\sf (2)} the centralized execution of automation and state changes. 
Similar to the {\em platform state machine} described in Section~\ref{sec:endorsement-check}, HomeAssistant has a state machine component that saves and updates all the devices and their states within HomeAssistant. 
In fact, all state changes to the devices occur through the state machine. 
\tool modifies this state machine to keep track of the
most recent state changes and their timestamps. 
It also hooks into the state machine to intercept the incoming state change requests, to mediate all API accesses. %
Furthermore, HomeAssistant has a component known as an {\em Event Bus}, which announces and logs the addition/removal of devices, among other events.
\tool adds callbacks in the Event Bus and re-instantiates the endorsement policies in response to any change in device deployment.
Finally, \tool keeps track of device-connectivity using HomeAssistant's built-in mechanisms, and falls-back to the next most restrictive policy in case a device becomes unavailable at runtime.
Note that no such loss of connectivity was noticed throughout our evaluation.

\section{Evaluation}
\label{sec:eval}

Our evaluation is guided by the following research questions:

\begin{enumerate}[label=$\bullet$ \textbf{RQ$_\arabic*$:},ref=\textbf{RQ$_\arabic*$},wide, labelwidth=!, labelindent=0pt]
\item \label{rq:feasibility} ({\em Feasibility of the policy model}.) Is it feasible to specify endorsement policies using only a small subset of trusted {\em endorsement attributes}, \ie only the read-only and designated device attributes? 
\item \label{rq:generalizability} ({\em Generalizability of the policy model}) Do feasible endorsement policies exist for endorsing AHOs other than \home?
\item \label{rq:security} ({\em Security.}) Does \tool prevent an attacker from escalating privilege to a high-integrity device (e.g., a camera) using one or more shared objects?
\item \label{rq:denials} ({\em False Denials.}) What is the rate of false denials in typical benign usage, i.e., when users intentionally cause AHO changes? 
\item \label{rq:performance} ({\em Runtime Performance}.) What is the performance overhead introduced by \tool?
\item \label{rq:integration} ({\em Integration Cost.}) How much effort is required to integrate and deploy \tool?
\end{enumerate}

We evaluate \ref{rq:feasibility}$\rightarrow$\ref{rq:integration} using a diverse array of empirical and experimental methods.
We first derive 1023 endorsement policies (\ie templates) for endorsing the \home AHO using the approach described in Section~\ref{sec:policy-spec}, which demonstrates that the proposed approach is feasible with even a limited endorsement attributes (\ref{rq:feasibility}).
Further, we demonstrate that \tool may generally apply to 5 additional AHOs drawn from platform documentation and prior work, as these AHOs may be endorsed using policies composed of $\approx$~5 device attributes on average~(\ref{rq:generalizability}).
We then deploy 11 devices (7 real, 4 virtual) in a real home, and perform a case study wherein we assume that the attacker's goal is to disable the security camera by tampering with state variables that the camera depends on.
Our experiments demonstrate that \tool correctly endorses the two AHOs that can be tampered with to disable the camera, thus demonstrating its effectiveness (\ref{rq:security}).

We perform 10 additional {\em realistic} scenarios obtained from (or inspired by) prior work~\cite{jm17}, which represent the various logical ways in which the user would intentionally change the AHOs endorsed in the case study.
Our evaluation demonstrates that \tool correctly allows all of the intentional AHO changes, resulting in no false denials (\ref{rq:denials}). 
We also measure the performance overhead incurred by \tools hooks using macro and microbenchmarks, demonstrating practical performance overheads (\ref{rq:performance}).

Finally, we evaluate the effort required to configure~\tool and integrate it into other platforms.
Particularly, we demonstrate that policy specification is a feasible, one-time effort performed by security experts, and {\em \tool automatically instantiates policies in the context of the user's home} (\ie as described in Sec.~\ref{sec:policy-spec}), drastically minimizing user effort.
Moreover, we distill the key properties that a platform would need to exhibit in order to integrate \tool,  qualitatively analyze 4 popular smart home platforms, and demonstrate that it is feasible to integrate \tool  with {\em modest engineering effort}~ (\ref{rq:integration}).

\begin{figure}[t]
    \centering
    \includegraphics[height=1.5in]{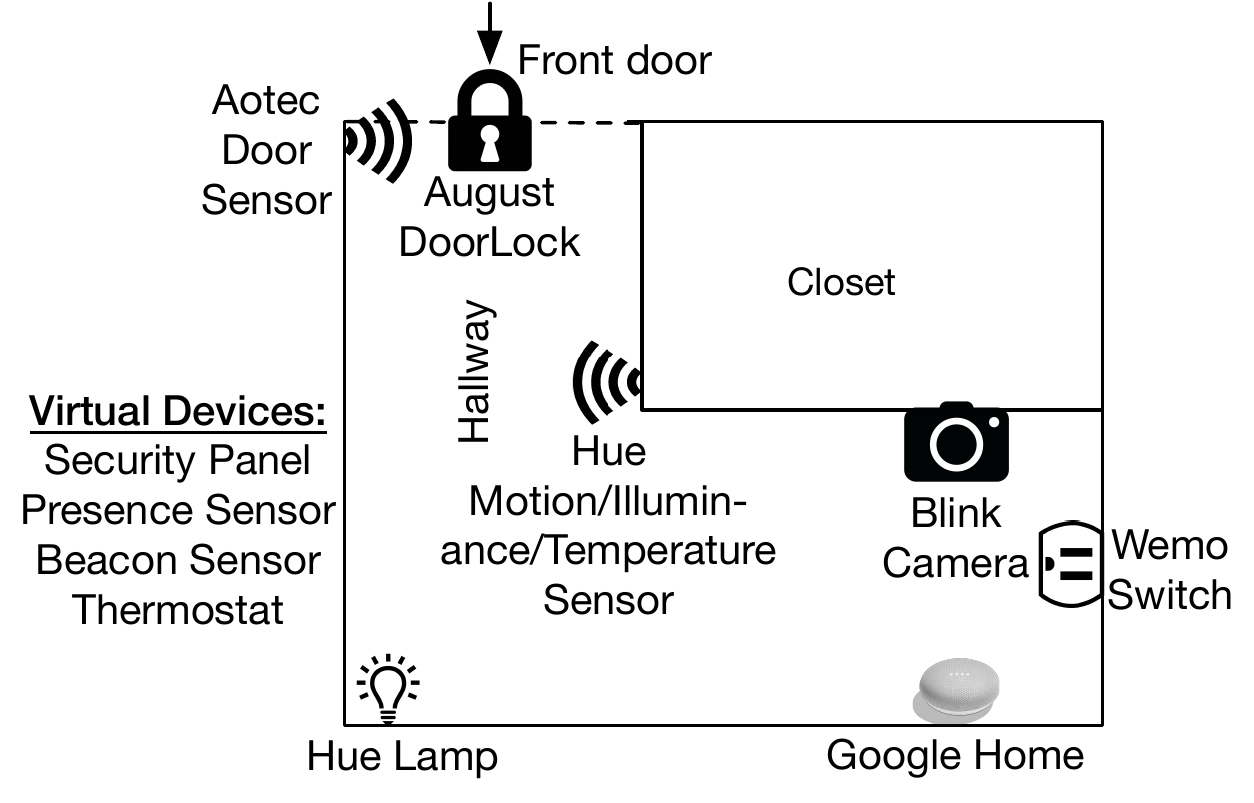} 
    \vspace{-1.5em}
    \caption{{\small Layout of the physical device placement}} \vspace{-1.5em}
\label{fig:device_placement}
\end{figure}

\myparagraph{Experimental Setup} We integrated \tool into HomeAssistant (v0.112.0.dev0), running on a Macbook Pro machine with 16GB of RAM.
Further, we connected 7 smart home devices to our smart home, and created 4 other {\em virtual} devices (see Table~\ref{tab:devices} in Appendix~\ref{app:eval-data} for full list).
Our choice of devices was limited by devices compatible with HomeAssistant, with a bias towards known/popular brands, as well as device types that would allow us to evaluate \tools endorsement policies.
We installed these devices in a room as shown in Figure~\ref{fig:device_placement}, influenced by effective deployment recommendations from prior work~\cite{jm17}.
To elaborate, we placed the August door lock and the Aotec door sensor on the main door, and the motion sensor in the hallway directly past the door. 
The motion sensor was placed such that it would not be triggered by the opening or closing of the door but would be triggered if someone walked down the hallway to get into the room. 
The Blink camera was placed inside the room. 
We interacted with the real devices physically and with the virtual devices through the HomeAssistant UI. 

\begin{table}
\centering
\scriptsize
\caption{{\small Sample policies for endorsing \home AHO (``away''$\rightarrow$``home'')}}
\vspace{-1.7em}
\setlength{\tabcolsep}{2pt}
\begin{tabular}{c|c}
 \Xhline{2\arrayrulewidth}
\hline
 & \textbf{Policy (location independent)} \\
 \Xhline{2\arrayrulewidth}                                                                                                                                                             
$P_1$ & $<${\sf \scriptsize security-panel, disarmed}$>$ $\wedge$ $<${\sf \scriptsize  motion-sensor, active}$>$                                                                                       \\ \hline
$P_2$ & $<${\sf \scriptsize  Doorlock, unlocked}$>$ $\wedge$ $<${\sf \scriptsize  presence-sensor, active}$>$   $\wedge$ $<${\sf \scriptsize  beacon, active}$>$
	\\ \hline
$P_3$ & $<${\sf Garage-doorlock,unlocked}$>$ $\wedge$ $<${\sf beacon,active}$>$
\\ \hline
$P_4$ & $<${\sf Security-panel,disarmed}$>$ $\wedge$ $<${\sf thermostat,motion.active}$>$
\\ \hline
\end{tabular}

\label{tab:strong_policies}
\vspace{-2em} 
\end{table}
\subsection{Feasibility of the Policy Model (\ref{rq:feasibility})}
\label{sec:feasibility-eval}

We do not expect users to possess all possible device-combinations that may endorse a specific AHO.
Instead, our approach must enable the specification of a  large and diverse set of candidate policies (\ie templates), to increase the possibility of finding at least {\em one} policy that contains the limited set of devices the user may possess.
Thus, our approach would be feasible if such a set of policies is possible, given the small number of trusted endorsement attributes.

Using the approach specified in Section\ref{sec:policy-spec},  
we were able to specify 1023 {\em unique} policies for endorsing the \home AHO (\ie the change from ``away'' to ``home''), consisting of 
 5 device-attributes on average, with the largest containing 10. 
Table~\ref{tab:strong_policies} contains 5 example policies generated using \tool (our online appendix~\cite{onlineappendix} provides the full list).
These policies were constructed from 10 device-attribute pairs that provide some inference for the \home AHO  change, of which 3 provide particularly strong endorsement of the user entering the home, as they denote authenticated entry (\ie~the door lock being unlocked, the garage door opened, and the security panel disarmed).
As a result, {\em even if the user had one of these 3 devices, they would be able to endorse \home}; the remaining 7 devices would only make the endorsement stronger.
Further, even if we include the policies that only have only 1 of these 3 device-attribute pairs (and other device-attributes), we get a total of 255 policies.
For example, as shown in Table~\ref{tab:strong_policies}, a user who has a door lock and motion sensor, or another who has a security panel, or another who has a garage doorlock and a presence sensor, would all be able to endorse \home using \tool.
This demonstrates that our approach is feasible, i.e.,  we can define a large number of diverse policies for an AHO (\ie\ \home), using a limited set of endorsement attributes, and hence, support a diverse array of device-deployment scenarios (\ref{rq:feasibility}).

\add{
\subsection{Generalizability of the Policy Model (\ref{rq:generalizability})}
\label{sec:generalizability}

To demonstrate the generalizability of our policy model, we consider 5 additional AHOs (\ie\ \secstate, \fire, {\tt water leak}, \illuminance, and \safety).
For each AHO, we identified endorsement attributes from the device-attribute map and then generated inferences using the process described in Section~\ref{sec:policy-spec}.
Our process resulted in 41 inferences (cumulatively) that would be useful for endorsement, with each AHOs being endorsed using {\em at least} 3 device-attributes (examples provided in Table~\ref{tab:inference} in Section~\ref{sec:policy-spec}).
This demonstrates that our approach is generalizable to AHOs other than \home, \ie similar policies are feasible for 5 other AHOs, as demonstrated by the substantial availability of applicable and trusted device-attributes (\ref{rq:generalizability}).

}

\subsection{Case Study: Protecting the Security Camera from Privilege Escalation (\ref{rq:security})}
\label{sec:eval-security}

\tool enables the following guarantee: given the attacker's goal to tamper with a high-integrity device that they cannot directly access (e.g., a security camera), \tools endorsement checks prevent the attacker from affecting malicious changes to {\em any AHO} that the aforementioned device depends on.
We demonstrate \tools effectiveness, with respect to the guarantee, using the attack described in the motivating example (Section~\ref{sec:motivating-example}).
As the camera depends on both the \home and \secstate AHOs, we experimentally demonstrate \tools effectiveness at preventing malicious modifications to both, using two scenarios:

\myparagraph{Malicious Scenario 1 -- Bob modifies \home}
We deploy a malicious third-party service controlled by the attacker, Bob. %
We assume that Alice has granted to the service the permission (\ie a REST API token) to write to the \home AHO. 
When Alice is out of the home, Bob writes to the value ``home'' to \home, to disable the camera.
Without \tool, the \home AHO will change, allowing Bob to remotely disable the security camera; however, we consider that Alice uses \tool with the policy $P_{\home_{1}}$({\footnotesize home}) enabled:
\begin{center}
{\small
$P_{\home_{1}}$({\footnotesize home}) $=$ (door-lock\_{lock} $==$ {\small\tt UNLOCKED} $\wedge$\\
			  motion\_{sensor} $==$ {\small \tt ACTIVE})$_{front-door}$
}
\end{center}
Thus, when Bob writes to \home, the policy $P_{\home_{1}}$({\footnotesize home}) is checked as follows:
\tool queries the state machine, and obtains the most recent change in the state of the door lock as well as the entryway motion detector.
Since the door lock was not unlocked, and the motion detector has also not detected motion recently, the policy returns a {\sf \small DENY} decision, and Bob's access is denied, preventing the attack.
It is also important to note that Bob could attempt to circumvent \tools policy by satisfying one of the two conditions in it, \eg by sliding a thin object (e.g., a card) through the door to trigger the motion sensor; however, the conjunction among device-attributes prevents this attack variation.

\myparagraph{Malicious Scenario 2 -- Bob modifies \secstate}
We deploy a malicious third-party service controlled by Bob, to which Alice has granted the permission to write to the \secstate AHO. 
Bob will attempt to set the \secstate to ``ok'' (as opposed to ``deter''), which will trigger a routine that turns off the camera. 
Without \tool, the \secstate state will successfully change, allowing Bob to disable the camera; however, we consider that Alice uses \tool with the policy $P_{\secstate_{1}}$({\footnotesize ok}) enabled:

\begin{center}
{\small
$P_{\secstate_{1}}$({\footnotesize ok}) $=$ (security-panel $==$ {\small\tt DISARMED} $\wedge$\\
			  motion\_{sensor} $==$ {\small \tt ACTIVE})$_{front-door}$
}
\end{center}

When Bob writes to \secstate, $P_{\secstate_{1}}$({\footnotesize ok}) is checked. 
Since the security panel wasn't disarmed and the motion sensor wasn't active recently, the policy returns a {\sf \small DENY} decision, preventing the attack.
To summarize, \tool successfully endorses both of the AHOs on the transitive attack path to the security camera, thereby preventing Bob from disabling the camera via a lateral privilege escalation.

\subsection{Endorsing Intentional State Changes (\ref{rq:denials})}
\label{sec:eval-intentional-changes}

We experimentally evaluate the behavior of \tool when users intentionally request state changes.
Users may themselves choose to change AHOs (e.g., by manually selecting \home mode in their platform mobile app after getting home), or by using a third-party service that automatically requests appropriate state changes based on the user's location (\eg when the user arrives home).
We derive a set of 10 {\em realistic user behavior scenarios} from prior work~\cite{jm17} to use as our benign examples, and enact those scenarios in our apartment testbed.
Note that \tool can provide endorsement policies to endorse all 10 scenarios, even when they contain a diverse array of devices, as previously demonstrated in Section~\ref{sec:feasibility-eval}.
Due to space constraints, we summarize 5/10 scenarios that are exemplary of \tools decisions in response to benign user behavior, with the rest available in Table~\ref{tab:scenarios} in Appendix~\ref{app:scenarios}.

\myparagraph{Scenario 1 -- Entering the house} Alice returns home, unlocks the front door lock and opens the door. 
She then closes the door and enters, triggering the motion sensor near the door.
Around the same time, her home/away tracker service requests a state change from away to home using REST API. 
In response to the request, \tool gathers the recent states of the devices to check against the active policy (\ie\ $P_{\home_{2}}$({\footnotesize home})).
\begin{center}
{\small
$P_{\home_{2}}$({\footnotesize home}) $=$ (door-lock\_{lock} $==$ {\small\tt UNLOCKED} $\wedge$\\
			  motion\_{sensor} $==$ {\small \tt ACTIVE} $\wedge$ door\_sensor $==$ {\small \tt ACTIVE})$_{front-door}$
}
\end{center}

Since the door lock was manually unlocked and the sensors were active recently, the policy is satisfied and \tool allows the change, as Alice intended.

\myparagraph{Scenario 2 -- Unlocking the house, and then leaving} Alice returns home and opens the front door after unlocking the front door lock. 
However, she gets a call from her office and leaves immediately without entering the house, accidentally also leaving the door open in the process.
Regardless, Alice's home/away service accidentally requests the \home AHO to change to ``home'' (\ie even when Alice has actually left), which would disable the camera and leave the home in an unmonitored state.

In response to the request, \tool gathers the recent states of the devices to check against the active policy (\ie\ $P_{\home_{2}}$({\footnotesize home})).
The policy constraints are {\em partially} satisfied, as the door lock was recently manually unlocked, and the door sensor's state changed to ``open''. 
However, as Alice did not enter, the motion sensor did not detect any motion, and the policy results in a {\em correct denial}, preventing an unsafe situation in which the camera is turned off while the home is vulnerable (i.e., the door is unlocked).
That is, \tools composite policy design leverages additional devices such as the motion detector to provide {\em stronger} endorsement, and hence, is useful for preventing accidental but unsafe changes.

\myparagraph{Scenario 3 -- A home without a smart door lock} Alice does not possess a smart door lock, but enables the following policy based on the devices in her home:
\begin{center}
{\small
$P_{\home_{3}}$({\footnotesize home}) $=$ (presence\_{sensor} $==$ {\small\tt ACTIVE} $\wedge$\\
			  motion\_{sensor} $==$ {\small \tt ACTIVE})$_{front-door}$
}
\end{center}
Alice comes home and enters through the front door, triggering the motion sensor and the presence sensor in the hallway       . 
At the same time, her service requests a change to \home.
In response, HomeEndorser gathers the recent device states to check against the active policy, $P_{\home_{3}}$({\footnotesize home}). 
Since both the sensors were recently active, the policy is satisfied and the state change is correctly allowed.
That is, \tools policy structure enables us to specify a variety of endorsement policies containing a diverse array of devices, which supports several home configurations, and prevents false denials due to the absence of certain devices.

\myparagraph{Scenario 4 -- Disarming the security panel and entering} 
In this scenario, Alice returns home and inputs the key-code in the security panel near the door, disarming the home. She then enters the home triggering the motion sensor. 
At the same time, a home security service requests change to the \secstate object on Alice's behalf, from ``deter'' to ``ok'', which if allowed, would disable the security camera, as well as any other security devices (\eg alarms).

HomeEndorser gathers the recent states of the devices to check against the active policy, $P_{\secstate_{1}}$({\footnotesize ok}) (provided previously in Section~\ref{sec:eval-security}). 
Since the security panel was manually disarmed and the motion sensor was active recently as well, the policy is satisfied and the state change is correctly allowed.

\myparagraph{Scenario 5 -- Direct state change request} Alice returns home, and manually changes the \home AHO to ``home'' using the HomeAssistant dashboard. 
HomeEndorser identifies that the request was not made through the REST API, and correctly allows it without checking the policy, thereby honoring the user's explicit intent.

Our evaluation demonstrates that \tool does not cause false denials under benign behavior.
That is, it generally does not deny state changes intended by the user, and when it does, its denials prevent {\em accidental} and harmful state changes by users (\ie and hence, are correct) (\ref{rq:denials}).
We further observed that in several cases, by the time the endorsement was requested (\ie time of REST API call), the relevant devices (\eg motion sensors) had reverted to their default states (\eg the motion sensor reverted to the ``inactive'' state).
This means that if we had only checked the {\em current} state of the devices, \ie the state at the time of endorsement, the endorsement check would have resulted in a {\em false denial}.
However, \tools approach of checking the {\em most recent} change prevents such potential false denials. 
Finally, we did not observe any device or automation crashes/failures in the home in response to \tools endorsement (see Appendix~\ref{app:compat} for details).

\begin{table*}
\centering
\footnotesize
\caption{{\small Performance overhead of \tool (in comparison with the unmodified HomeAssistant baseline)}}
\label{tab:perf_table}
\vspace{-1.5em}
\setlength{\tabcolsep}{2pt}
\begin{tabular}{c|l|c|c|c|c}
 \Xhline{2\arrayrulewidth}
\hline
\textbf{No.} & \textbf{Operation} & \textbf{HomeAssistant Baseline (ms)} & \textbf{\tool (ms)} & \textbf{Overhead(ms)} & \textbf{Overhead(\%)}\\
 \Xhline{2\arrayrulewidth}                                                                                                                                                             
1. & Policy Instantiation ({\em Boot up time}) & 23.851 $\pm$ 1.738 & 33.669 $\pm$ 5.042 & 9.818 & 41.16                                                                                        \\ \hline
2. & Policy update during runtime & - & 4.350 $\pm$ 0.515 & - &  -
\\ \hline
3. & Changing non-endorsed AHO ({\em Hook invocation cost}) & 9.854 $\pm$ 0.723 & 9.916 $\pm$ 0.814 & 0.062 &  0.63
\\ \hline
4. & Changing endorsed AHO ({\em Endorsement check cost}) & 9.451 $\pm$ 0.605 & 10.367 $\pm$ 0.482 & 0.916 & 9.69
\\  \Xhline{1.5\arrayrulewidth} 

5. & Automation execution with endorsed AHO & 16.582 $\pm$ 2.388 & 18.598 $\pm$ 0.669 & 2.016 & 12.16
	\\ \hline
6. & Automation execution with non-endorsed AHO & 14.609 $\pm$ 1.026 & 14.311 $\pm$ 0.477 & -0.298 & -2.04 
\\ \hline

\end{tabular}

\vspace{-1.5em}
\end{table*}
\subsection{Runtime Performance (\ref{rq:performance})}
We measure the performance of \tool using a combination of macro and microbenchmarks (\ref{rq:performance}).

\myparagraph{Methodology} We compute microbenchmarks to capture each aspect of the platform that \tool affects, in particular, the time taken for {\sf (1)}~{\em policy instantiation} (\ie additional delay at boot time), {\sf (2)} {\em policy update} during runtime (\ie overhead of device addition/removal) {\sf (3)} the {\em endorsement hook invocation cost} (\ie overhead of an API call to a state {\em not being endorsed}), and, {\sf (4)} the {\em endorsement check} (\ie overhead of an API call to a state being endorsed).
Furthermore, we perform 2 macrobenchmarks to assess the overall end-to-end impact of \tool on the execution times of remote IoT services that leverage the REST API for {\sf (5)} executing an automation involving an AHO being endorsed, and {\sf (6)} executing an automation involving an AHO not being endorsed.%
We measure the worst-case performance by performing all measurements with the largest policy, i.e., the policy involving the maximum number (\ie 7) of device-attributes that can be instantiated in our device deployment (\ie $P_{109}$, online appendix~\cite{onlineappendix}). 
We compute baselines for the benchmarks using an unmodified build of HomeAssistant with the same devices. We perform each experiment 50 times in both environments \ie 50 executions each on HomeAssistant with \tool and HomeAssistant without \tool. Table~\ref{tab:perf_table} shows the mean results with 95\% confidence intervals.

\myparagraph{Results} 
As seen in Table~\ref{tab:perf_table}, \tool has negligible performance overhead for operations that do not involve the AHO being endorsed, \ie the hook invocation cost when calling an API that modifies a non-endorsed AHO (microbenchmark), as well as executing an automation that changes a non-endorsed AHO (macrobenchmark), with the overhead falling within the error margin (which explains the negative overhead for operation 6).

For endorsed AHOs, \tool adds only 0.916ms (9.69\% overhead) to an AHO-change invoked via an API call (microbenchmark), and adds 2.016ms (12.16\% overhead) to the overall execution time of an automation execution that changes an endorsed AHO (macrobenchmark). 
In fact, the maximum overhead of 9.818ms (41.16\%) that \tool adds is to the overall bootup time of HomeAssistant, which is not that frequent, and not perceivable by the user. 
This is primarily due to the fact \tools Policy generator is called repeatedly as HomeAssistant adds devices during bootup. 
After the bootup, the overhead to update policies when devices get added or removed is only 4.350 ms.  
Finally, we note that the endorsement check overhead may not be dependent on the policy size, as we have confirmed via inspecting HomeAssistant's code that its state machine obtains device state changes in parallel.

\subsection{Effort Required to Integrate and Configure \tool (\ref{rq:integration})}
\label{sec:integration}

This section describes the effort that may be required to deploy and integrate \tool.
Particularly, we demonstrate that {\sf (1)} experts may specify a large number of endorsement policies for \tool using the systematic semi-automated methodology defined in Section~\ref{sec:policy-spec} with minimal effort and time, {\sf (2)} \tool can {\em automatically} instantiate its policies in end-user homes as per the availability and placement of devices, which significantly reduces configuration effort, and finally {\sf (3)} most platforms may directly integrate \tool without any design-level extensions, which indicates its practicality, and promise of future deployment (\ref{rq:integration}).

\myparagraph{1. Policy specification} \tools policies are specified by experts, and automatically instantiated in the end-user's home based on the devices deployed at various locations.
The semi-automated, data-driven, process for specifying policies is a one-time effort, as described in Section~\ref{sec:policy-spec}, and policies only need to be updated when new functionality emerges in an entire category of devices such as door locks (\ie and not individual brands), or when an entirely new category of device is introduced to the market.
To generate the 1023 policies for endorsing \home, as well as the 510 device-attributes that could help endorse 5 other states, it took 2 authors around 4 workdays, which is feasible given that the policies will provide endorsement in user homes with diverse device configurations.%

\myparagraph{2. Deployment in the User's Home} The user deploys \tool by simply deploying the smart home platform that it is integrated into (\eg HomeAssistant, as implemented in the paper). 
The deployed platform comes with pre-specified policies built-in, which are instantiated in the context of the user's device setup (as described in Section~\ref{sec:endorsement-check}).
Most platforms already require the user to provide device-location for functionality, and \tool simply piggy-backs on this information to instantiate policies.
The only additional effort on the user's part is selecting the AHOs that they would like to endorse.
Finally, \tool seamlessly updates the policy configuration with minimal performance overhead when new devices are added or existing devices removed.

\myparagraph{3. Platform integration} Through lessons learned during our implementation of \tool in HomeAssistant, we distill 4 key platform-properties that are necessary to successfully integrate \tool in a smart home platform, in a way that satisfies the general purpose of physical home endorsement, as well as certain enforcement-related design goals (\ref{goal:complete-mediation}$\rightarrow$\ref{goal:freshness}).
We then discuss these identified properties in the context of 4 popular platforms, namely  IoTivity~\cite{iotivity}, OpenHAB~\cite{openhab}, \smartthings~\cite{smartthingsv3}, and \nest/Google Assistant~\cite{wwGoogleAssistant}, and estimate the effort that would be required for the wide-spread integration of physical home endorsement, across all platforms.
We now describe the properties, followed by our qualitative analysis of platforms in terms of ease of integration.

\emparagraph{Property 1 ($Prop_1$) - Ability to obtain device states} In order for \tool to correctly enforce the policies, it has to be able to obtain states from all devices involved in the policies. Ideally, the platform should have a Platform State Machine that can readily provide device state information. This property is necessary for the general correctness of enforcement, and no design goal in particular.

\emparagraph{Property 2 ($Prop_2$) - Complete mediation of state changes} For \tool to intercept all incoming API requests that seek to change AHOs, and trigger endorsement checks (\ref{goal:complete-mediation}), the platform must have a central component that intercepts all the API requests. Furthermore, this component must be unmodifiable by third parties (\ref{goal:tamperproofness}).

\emparagraph{Property 3 ($Prop_3$) - Timestamp information of device states} \tool requires {\em fresh} devices state information in order to remove any false positives that can occur because of devices reporting cached states or the platform itself reporting the last known state as the current state because of a non-responsive device (\ref{goal:freshness}).

\emparagraph{Property 4 ($Prop_4$) - Ability to monitor device changes} \tool needs the ability to dynamically adapt its policies based on the current setup of the smart home, and hence, the platform needs to monitor the addition, removal, and change in placement of devices.

\begin{table}
\centering
\footnotesize
\caption{{\small The (minimal) cost of Integrating \tool with respect to the properties identified in Section~\ref{sec:integration}}}
\vspace{-1em}
\setlength{\tabcolsep}{2pt}
\begin{tabular}{c|c|c|c|c|c}
 \Xhline{2\arrayrulewidth}
\hline
 & \textbf{HomeAssistant} & \textbf{IoTivity} & \textbf{OpenHAB} & \textbf{SmartThings} & \textbf{\nest} \\
 \Xhline{2\arrayrulewidth}                                                                                                                                                             
$Prop_1$ & \ye & \ye & \ye & \pa &  \pa                                                                                     \\ \hline
$Prop_2$ & \ye & \ye & \ye & \ye & \pa     
	\\ \hline
$Prop_3$ & \ye & \no & \ye & \ye & \pa     
\\ \hline
$Prop_4$ & \ye & \ye & \ye & \pa & \pa     
\\ \hline
\end{tabular}
\begin{flushleft}
{\footnotesize \ye{} = Directly portable, \pa{} = Directly portable, but needs confirmation from source code, \no = design-level constraint/extension
}
\end{flushleft}
\label{tab:feasibility_study}
\vspace{-1.5em} 
\end{table}

\add{
Table~\ref{tab:feasibility_study} shows the results of our platform analysis based on $Prop_1\rightarrow~Prop_4$, and particularly, demonstrates that only IoTivity requires a design-level extension for integrating \tool  (in terms of $Prop_3$), and {\em all other platforms may feasibly integrate \tool}, sometimes with negligible engineering efforts. 
Each integration may require minor adjustments, \eg for HomeAssistant, we modified the state machine for complete mediation, whereas in OpenHAB, we would need to implement a reference monitor across various bindings (\ie hook into services exposed as bindings), in a manner similar to how extensible access control has been previously implemented in OSes such as Linux~\cite{wcs+02} and Android~\cite{hnes14}.
While IoTivity provisions states to the reference monitor ($Prop_2$), a state machine with the ability to collect freshness information will need to be implemented ($Prop_3$), which would be a design-level extension.

We mark certain properties for \nest and \smartthings as \pa\ in Table~\ref{tab:feasibility_study} because while we can almost certainly say that they possess the required properties to integrate \tool with minimal engineering efforts, we would need source code to confirm.
\nest is a particularly interesting example, because while it does exhibit all of the properties, making it an ideal platform to integrate \tool into, it currently supports $Prop_1$ for only a limited set of devices, \ie the \nest-owned devices that are in its data store (\eg the \nest camera, \nest smoke alarm). 
However, this may be easily remedied by allowing third-party devices to be integrated into the data store as well, so that they can report their states for the platform to access.
Further, while SmartThings does not protect AHOs (\eg\ \home) with permissions, it does provide a permission framework that mediates most similar accesses (and hence, has the ability to protect AHOs).
Hence, integrating \tool would simply require the placement of endorsement checks alongside the permission enforcement, which is engineering effort similar to what we observe for OpenHAB in terms of $Prop_2$. 
Furthermore, as an app can be configured to support multiple devices and observe their states ($Prop_1$) and as it can also obtain the state change timestamps from the platform logs ($Prop_4$), we can say that \smartthings may be able to perform these actions at the platform-level.

}

\section{Related Works}
\label{sec:related_works}

Much prior work in smart home security has focused on reducing
the misuse of sensitive APIs, through improvements to the platform
permission model (\eg risk-based permissions~\cite{rfep18},
functionality-based enforcement~\cite{lck+17}), providing users with
context of an API access (\eg ContexIoT~\cite{jcw+17}),
enforcing least privilege using context from the app's description
(\eg SmartAuth~\cite{tzl+17}), and, when all else fails, forensic
analysis by integrating fine-grained provenance tracking into apps
(\eg ProvThings~\cite{whbg18}).  At first glance, our problem may also
seem to be one of over-privilege or API misuse; however, we assert
that addressing this problem as API misuse by retrofitting existing
defenses would be both ineffective and impractical.  \add{The
approach of removing privileges (or completely preventing
third-parties from accessing AHOs) would be impractical because
several third-party integrations may truly need access to AHOs, and
denying access may come at prohibitive usability costs, as we
previously discussed in Section~\ref{sec:need-checks}.  Instead, \tool
presents a more direct solution to the lack of integrity for AHOs in
the smart home.}

An alternate approach to securing shared states (including AHOs) is
centralizing them and only allowing trusted third parties to modify
them, as Schuster et al. explored using ``environmental situation
oracles''~\cite{sst+18} or ESOs.
The ESO model is mainly geared towards providing {\em privacy} (and
not integrity), \ie to prevent several parties from persistently
accessing the user's private information (\eg location) to generate a
shared state (\eg\ \home), by only allowing one dedicated trusted app
per shared state.  The key difference between ESOs and \tools
endorsement is that our approach does not require a trusted party to
take upon itself the responsibility of accurately generating the value
for a shared state, \add{which may introduce compatibility issues by requiring a
single app for every state to be approved by various stakeholders such as users, platform vendors, and device vendors.  
Instead,
\tool applies an endorsement policy that can {\em sanity check} a
proposed AHO modification based on consistency with expected device states.}

Prior work has also studied transitive problems occurring in the smart
home, mostly in the context of {\em app
  chaining/interactions}~\cite{sab+17}, where the premise is that two
or more applications could accidentally (or maliciously) trigger one
another, putting the system in an overall unsafe
state~\cite{sab+17,csk+20,ctm19,cmt18,nsq+18}.  Prior defenses against
app-chaining attacks involve static analysis (\eg
Soteria~\cite{cmt18}, IoTSAN~\cite{nsq+18}), or runtime rule-based
enforcement (\eg IoTGuard~\cite{ctm19}).  \add{Further, Ding and Hu
  demonstrated that such chains can also exploit the physical
  interactions among devices in the home~\cite{dh18} and build the
  IoTSafe system to predict application interactions at
  runtime~\cite{dhc21}.}  \tool differs from prior work on app chains
(whether or not they involve physical interactions) in two key ways.
\add{First, prior defenses are geared towards preventing attacks that
  involve two apps (\ie app interaction), and {\em are not designed to
    defend against arbitrary API calls that modify high-integrity
    AHOs}, which is a complementary research goal.  Second, prior work
  relies on the ability to view and/or instrument apps, whereas \tool
  is app-agnostic, only relying on local observations of device
  states.  This is a prudent decision, as most popular platforms no
  longer host third-party services, as instead such services are
  hosted from their own cloud environments and access the users'
  platforms via platform APIs.}

\tool relies on trusted {\em endorsement attributes}, \ie attributes
that cannot be modified by the adversary through API calls, but \tool
does not account for byzantine faults in devices.  However, there is
prior work in the area of validating states reported by devices,
particularly the recent Peeves system by Birnbach et al.~\cite{bem19}.
Peeves generates fingerprints of device events, in the form of
physical changes that they {\em cause}, which are sensed by other
trusted sensors, which attest to the validity of the reported device
state.  \add{However, note that \tool
and Peeves validate orthogonal types of states in the home, which lie
at different levels of abstraction, and hence, have distinct design
goals.  
That is, while Peeves focuses on sensor fingerprint accuracy
and precision, \tool focuses on building expressive policies that
involve a variety of device states to ensure compatibility with most
end-user deployments (\ref{goal:policy}), and on realizing key
reference monitor properties (\ref{goal:complete-mediation},
\ref{goal:tamperproofness}, \ref{goal:freshness}).
The device states that Peeves validates form the building
block of \tools endorsement policies that endorse {\em higher-level},
{\em abstract} home attributes (\eg\ \home, \secstate).  Hence, \tool
will benefit from the additional robustness and fault tolerance
enabled by complementary approaches such as Peeves, although neither
system promises byzantine fault tolerance.}

Finally, a related problem of endorsing operations has also been
explored in Android.  Android devices contain many sensors (e.g.,
camera and microphone) that apps may use if authorized.  Android
depends on users to authorize the sensors that apps may use, but a
malicious app may access a sensor at will once
authorized~\cite{comm_surv21}.  To address this problem, researchers
have explored methods to endorse authorized sensor operations by
comparing the context of the sensor operation request to authorized
contexts.  For instance, user-driven~\cite{rkm+12,ringer16ccs} access
control requires that applications use system-defined GUI gadgets
associated with particular operations
to endorse the sensor operation associated with a user input event
unambiguously.  The AWare~\cite{petracca17usenix} and EnTrust
~\cite{petracca19usenix} systems permit applications to choose their
own GUI gadgets, but restrict applications to employ the same GUI
configuration (e.g, layout) to endorse a sensor operation.
For IoT systems, rather than GUI contexts, we find that state changes are
associated with physical events, enabling them to be endorsed based on
properties of those physical events.

\section{Threats to Validity}
\label{sec:threats-to-validity}

\add{

With \tool, we seek to lay the groundwork for secure and practical endorsement for abstract home objects in the smart home, through our systematic and data-driven approach. 
However, there are certain threats to the validity of our evaluation and results, as we discuss below:

\myparagraph{1. Byzantine Fault Tolerance} We rely on devices to not be compromised and to report correct states, as stated in Section~\ref{sec:threat_model} (although \tool does dynamically adapt to devices that may be offline/non-responsive). 
That said, while \tool does not account for byzantine fault tolerance, it is complementary to prior efforts~\cite{bem19} that aim to validate device states via fingerprinting, and combining the two approaches is a promising future direction.

\myparagraph{2. Completeness and Rigor of Policy Specification} The device-attribute map used for identifying endorsement attributes was automatically constructed from platform-provided resources (Section~\ref{sec:policy-spec}). This map, consisting of 510 device-attribute-pairs, is an {\em evolving dataset} that is {\em as complete as the information sources} used to derive it (\eg device handlers, capability maps), and new device attributes/types that emerge may be integrated into it with minimal effort.
Furthermore, we used a systematic, 2-author, open coding approach to identify the endorsement attributes from this map that would enable endorsement of each of the 6 AHOs (Section~\ref{sec:implementation}).
This approach resulted in negligible disagreements (\ie only 2.4\%), illustrating high confidence in the identified inferences (our online appendix~\cite{onlineappendix} provides the complete dataset).
We note that our expert-defined policies are {\em only meant to be a seed-list}, which can be expanded upon with community support.

\myparagraph{3. Multi-user smart homes} Similar to most prior work in the area of smart home API misuse~\cite{jcw+17,rfep18,lck+17,whbg18}, \tool does not claim to address multi-user scenarios, which are a novel but orthogonal design challenge.  
Further, we note that certain AHOs (\eg fire) may be independent of the number of users, and moreover, \tools location-specific policy-predicates (Section~\ref{sec:policy-design}) may also mitigate the implications of multiple simultaneous device-interactions, although we do not make formal claims.

\myparagraph{4. Device Availability and Placement} \tools policies (\ie at least the seed list specified in this paper) consist of a diverse range of devices. 
Hence, the user does not need to have all the devices, as HomeEndorser can automatically choose the most restrictive policy that is applicable to a user’s home based on device-availability and placement information obtained from the platform (Section~\ref{sec:endorsement-check}). 
However, as stated in Section~\ref{sec:policy-spec}, we assume optimal device placement and sensor configuration to be out of scope, and direct the reader to complementary related work that informs users on optimal deployment~\cite{jm17}. 
}

\section{Conclusion}
\label{sec:conc}

In this paper, we presented the \tool framework, which validates proposed changes in high integrity abstract home objects (AHOs) in correlation with device states.  
\tool aims to address the threat posed by compromised/malicious services that may use their authorized API access to change AHOs in a way that causes high-integrity devices to behave unsafely.  
To do this, \tool leverages the insight that that changes in AHOs are inherently tied to a home’s physical environment, enabling each sensitive change to be validated, or
{\em endorsed}, by correlating to changes in physical device states.
\tool provides a policy model for specifying endorsement policies
in terms of device state changes, 
and a platform reference monitor for endorsing all API requests to
change AHOs.  We evaluate \tool
on the HomeAssistant smart home platform, finding that we can feasibly derive policy rules for \tool to endorse changes to 6
key AHOs, preventing malice and accidents with feasible performance overhead.
Finally, we demonstrate that \tool is backwards compatible with most popular smart home platforms, and requires modest human effort to configure and deploy.

\bibliographystyle{acm-reference-format}
\bibliography{bib/bibliography,bib/onlinebibliography,bib/phone,bib/cryptoapi,bib/taint,bib/iot,bib/mutation-references,bib/misc,bib/mutation_security,bib/os}

\newpage
\appendix
\begin{figure}[t]
	\centering
    \includegraphics[bb=0 0 374 161, width=2.5in]{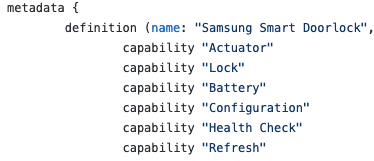}
    \vspace{-0.5em}
    \caption{\small Specification of device-attributes in the SmartThings device handler preamble.} 
    \label{fig:devicehandler}
    \vspace{-1.5em}
\end{figure}

\section{Device List for Evaluation}
\label{app:eval-data}
\begin{table}
\centering
\footnotesize
\caption{{\small Real and Virtual Devices in Evaluation}}
\begin{tabular}{c|c|c}
 \Xhline{2\arrayrulewidth}
\hline
\textbf{Device} & \textbf{real/virtual} & \textbf{Number} \\
 \Xhline{2\arrayrulewidth}                                                                                                                                                             
August Door Lock & $real$ & 1
\\ \hline
Blink Camera & $real$ & 1                                                                                      
\\ \hline
Philips Hue Motion+Illuminance+Temp. Sensor & $real$ & 1                                                                                      \\ \hline
Aotec Door Sensor & $real$ & 1
\\ \hline
Security Panel & $virtual$ & 1
\\ \hline
Presence Sensor & $virtual$ & 1                                                                                       
\\ \hline
Beacon Sensor & $virtual$ & 1                                                                                       
\\ \hline
Thermostat & $virtual$ & 1                                                                                       
\\ \hline
Wemo Switch & $real$ & 1                                                                                       
\\ \hline
Philips Hue Lamp & $real$ & 1                                                                                       
\\ \hline
Google Home & $real$ & 1                                                                                       
\\ \hline
\end{tabular}
\label{tab:devices}
\vspace{-1.5em} 
\end{table}
 
Table~\ref{tab:devices} shows the real as well as the virtual devices we integrated into HomeAssistant to conduct the experiments.

\section{Scenarios}
\label{app:scenarios}
\begin{table*}[t]
\centering
\scriptsize
\caption{{\small Realistic Scenarios used in evaluation of \tool}}
\label{tab:scenarios}
\begin{tabular}{p{1.5cm}|p{3.5cm}|p{2cm}|p{3.5cm}|p{4.5cm}}
\Xhline{2\arrayrulewidth}
\multicolumn{0}{c|}{\textbf{Scenario}}   & \multicolumn{0}{c|}{\textbf{User Behavior}}                                                                                                                                                                                                                                                                                                                                                                                                 & \multicolumn{0}{c|}{\textbf{State change request}} & \multicolumn{0}{c|}{\textbf{Policy involved}} & \multicolumn{0}{c}{\textbf{\tool behavior}}
\\ \Xhline{2\arrayrulewidth}
$S6:$ Entering the house without closing the door &  Alice comes home and opens the front door after unlocking the front door lock manually. She leaves the door open, and comes in by triggering the motion sensor in the hallway. & An app requests a state change from \away to \home using REST API. & 

{\small $P_{home_{2}} =$ (door-lock\_{lock} $==$ {\small\tt UNLOCKED} $\wedge$
			  motion\_{sensor} $==$ {\small \tt ACTIVE} $\wedge$ door\_sensor $==$ {\small \tt ACTIVE})$_{front-door}$ } 	
	  
 & \tool gathers the recent states of the devices. Since the door-lock was manually unlocked and the sensors were active recently, the policy is satisfied and the state change is {\em endorsed}. \\ \hline

$S7:$ Unlocking the home and leaving &  Alice comes home and opens the front door after unlocking the front door lock manually. She steps outside immediately after entering and closes the door behind her, thereby triggering the door sensor again. & An app requests a state change from \away to \home using REST API.  &

 same as {\small $P_{home_{2}}$}
 
 & \tool gathers the recent states of the devices. The door lock was manually unlocked and the door sensor was triggered both when opening and closing the door. However, the motion sensor was not active recently so the policy is not satisfied and the state change is {\em denied}. \\ \hline

$S8:$ Diverse device setup1 &  Alice comes home and opens the front door after unlocking the front door lock manually. She comes in by triggering the motion sensor and the beacon sensor in the hallway. & An app requests a state change from \away to \home using REST API. &

{\small $P_{home_{3}} =$ (door-lock\_{lock} $==$ {\small\tt UNLOCKED} $\wedge$
			  motion\_{sensor} $==$ {\small \tt ACTIVE} $\wedge$ door\_sensor $==$ {\small \tt ACTIVE} $\wedge$ beacon\_sensor $==$ {\small \tt ACTIVE})$_{front-door}$ }  	
 
 & \tool gathers the recent states of the devices. Since the door-lock was manually unlocked and the sensors were active recently, the policy is satisfied and the state change is {\em endorsed}. \\ \hline

$S9:$  Diverse device setup2 &  Alice comes home and inputs the keycode in the security panel near the door, disarming the home. She then enters the home triggering the motion sensor in the hallway.  & An app requests a state change from \away to \home using REST API. &

{\small $P_{home_{4}} =$ (security-panel $==$ {\small\tt DISARMED} $\wedge$ motion\_{sensor} $==$ {\small \tt ACTIVE} $\wedge$ presence\_{sensor} $==$ {\small \tt ACTIVE})$_{front-door}$} 	

 & \tool gathers the recent states of the devices. Since the security panel was manually disarmed and the motion sensor was active recently as well, the policy is satisfied and the state change is {\em endorsed}. \\ \hline

$S10:$ Home with just 2 devices &  Alice comes home and unlocks the front door lock manually. She comes in by triggering the motion sensor in the hallway. & An app requests a state change from \away to \home using REST API. &

{\small $P_{home_{5}} =$ (door-lock\_{lock} $==$ {\small\tt UNLOCKED} $\wedge$
			  motion\_{sensor} $==$ {\small \tt ACTIVE} } 	
 
 & \tool gathers the recent states of the devices. Since the door lock was manually unlocked and the motion sensor was active recently, the policy is satisfied and the state change is {\em endorsed}.
 \\ \Xhline{2\arrayrulewidth}                                                                                                                                       
\end{tabular}
\end{table*} 
Table~\ref{tab:scenarios} shows the remaining realistic scenarios (from Section~\ref{sec:eval-intentional-changes}).

\section{Backwards Compatibility} 
\label{app:compat}

One of our key objectives is to build a practical framework that does not crash or in any other way impede legitimate functionality.
Therefore, we evaluate the backwards compatibility of \tool with automation developed by real users or app developers (\ie IoT apps or user-driven routines).

\myparagraph{Methodology}
We randomly selected 20 automations from the SmartThings repository~\cite{smartapp-repo}. 
We then leveraged our existing deployment of \tool and set up 10 of these automations in HomeAssistant that were applicable to the devices in our deployment, filtering out the remaining 10 for which we did not have devices.
We then triggered the automations throughout the day, in a random order and at random times, while constantly monitoring for system crashes, dropped or delayed device actions, or any other unusual device, automation, or platform failures.
Note that since this is an automated experiment, certain automated triggers may be denied due to endorsement checks; we consider these denials as legitimate, and not failure cases.

\myparagraph{Results} 
We did not observe any crashes or device failures. 
Note that the selection of routines included one involving \home shared state changing from {\em away} to {\em home} that was denied throughout the day, \ie\ {\em when user is home, disarm the security panel}. 
On the other hand, routines that were changing states not endorsed by \tool were allowed throughout the day.

\end{document}